\documentclass[12pt]{article}
\usepackage[margin=1in]{geometry}
\usepackage{graphicx}
\usepackage{caption}
\usepackage{amsmath,amssymb}
\usepackage{bm}
\usepackage[table]{xcolor}
\usepackage[
colorlinks=true,
allcolors=black
]{hyperref}
\usepackage{nameref}
\usepackage{float}

\usepackage{lmodern}
\usepackage{threeparttable}
\usepackage{tablefootnote}
\usepackage{subfigure}
\usepackage{soul}

\definecolor{RedditColor}{RGB}{153,204,204}
\definecolor{TwitterCovid}{RGB}{204,230,255}
\definecolor{TwitterOther}{RGB}{204,255,230}

\newcommand{\bite}{\begin{itemize}}
\newcommand{\eite}{\end{itemize}}
\newcommand{\benu}{\begin{enumerate}}
\newcommand{\eenu}{\end{enumerate}}

\title{\bf Quantifying opinion homophily in online social networks: {A bounded confidence perspective}}

\author{Yangyang Luan\thanks{School of Mathematics and Statistics, Wuhan University, Wuhan 430072, China.}$\;\,$\thanks{Division of Automatic Control, Department of Electrical Engineering, Link\"oping University, SE-58183 Link\"oping, Sweden.} \and 
Camilla Ancona\thanks{Department of Electrical Engineering and Information Technology, Universit\`{a} degli Studi di Napoli Federico II, Napoli, 80125, Italy.} \and 
Carmela Bernardo\thanks{Department of Engineering, University of Sannio, 82100, Benevento, Italy.} \and 
Valentina Pansanella\thanks{Institute of Information Science and Technologies ``A. Faedo'' (ISTI), National Research Council (CNR), Via Giuseppe Moruzzi 1, Pisa, Italy.} \and 
Francesco Lo Iudice\footnotemark[3] \and 
Giulio Rossetti\footnotemark[5] 
\and 
Francesco Vasca\footnotemark[4] \and 
Xiaoqun Wu\thanks{College of Computer Science and Software Engineering, Shenzhen University, Shenzhen 518060, China.} \and
Claudio Altafini\footnotemark[2] \thanks{To whom correspondence should be addressed: Email: {\tt\small claudio.altafini@liu.se}}}
\date{}

\begin{document}
	
\maketitle

\begin{abstract}
The concept of homophily is pervasive in online social media. While many empirical studies have relied on external sociodemographic traits to investigate it, significantly less is known about homophily at the cognitive level, that is, at the level of shared opinions or values.
For such ``value homophily'', in this paper {we study interval-based patterns of opinion homophily from a bounded confidence perspective.}
We consider three heterogeneous datasets from Reddit and Twitter covering polarizing issues, with user opinions quantified via sentiment analysis and fact-checking, and analyze the interaction networks formed by weaker (reply-based) and stronger (follow-based) social ties.
Our findings show that users' {interaction neighborhoods are significantly more concentrated in opinion space} than expected by chance, with tie strength and issue polarization further amplifying this effect.
Moreover, users often exhibit asymmetric tolerance 
%thresholds, more readily accepting moderate, mainstream-aligned deviations than radical departures from their current ideological stance.
{ranges}, {with asymmetry typically directed toward locally mainstream positions rather than more radical or opposing ones.}
%These results provide evidence consistent with bounded-confidence-style selection as a plausible mechanism for online value homophily.
These findings {support a bounded confidence interpretation of} online value homophily.
\end{abstract}

\section*{Introduction}
\label{sec:Introduction}

Homophily, the tendency of individuals to associate preferentially with others who share similar attributes, is a cornerstone in understanding the structure and dynamics of social interactions and is well captured by the adage ``birds of a feather flock together''~\cite{mcpherson2001birds}. 
Many studies in the last decade have focused on investigating homophily in online social media \cite{khanam2023homophily}. 
Users on these platforms tend to form online communities that are homogeneous along certain attributes, which can range from demographic traits such as ethnicity, age, and gender, to social markers like education, occupation, interests, or cultural background.
Also the forms of association can be very diverse and can include friendship, shared interests, geographical proximity, etc.~\cite{mcpherson2001birds,khanam2023homophily}. 
The concept has been used to explain a wide range of sociological effects, from community formation~\cite{currarini2009economic} to segregation~\cite{leszczensky2015ethnic}, from innovation diffusion~\cite{dearing2018diffusion} to health behavior adoption~\cite{centola2011experimental}, from political discourse amplification~\cite{halberstam2016homophily} to fashion trends~\cite{zhang2018fashion}.

Homophily is particularly pervasive in the sphere of politically charged or ideologically sensitive online themes.
Social media networks often segregate into partisan communities, reinforcing like-minded interactions and limiting exposure to opposing viewpoints. For instance, Twitter follow graphs reveal that users cluster tightly along ideological lines \cite{barbera2015birds}, retweet networks form polarized left/right communities \cite{conover2011political,flamino2023political}, and news or content spreads preferentially among like-minded users \cite{halberstam2016homophily,brady2017emotion}. Such patterns contribute to the formation of so-called ``echo chambers'' (in which individuals are exposed only to information from like-minded individuals) and ``filter bubbles'' (in which algorithms serve content aligned with a user's prior behavior) \cite{bakshy2015exposure}.
Extensive prior research confirms that shared political views strongly drive online social ties, as it does with off-line ones \cite{barbera2015birds,huber2017political,bail2018exposure,cinelli2021echo,mosleh2021shared}.

An axis frequently used to classify the different forms of homophily is the distinction into ``attribute-based homophily'' associated with exterior traits (such as sociodemographic attributes~\cite{kim2011modeling}, interaction frequency, minority~\cite{karimi2018homophily} or segregation patterns~\cite{mayerhoffer2025social,mayerhoffer2025networks}), and ``value homophily'' associated with interior traits and fundamental beliefs (such as cognitive and psychological factors, core values, and opinions)~\cite{mcpherson2001birds,lazarsfeld1954friendship}.
Compared to attribute-based homophily, value homophily exerts a deeper influence on group identity, cooperation, and conflict dynamics \cite{khanam2023homophily,huber2017political,mosleh2021shared}.

Yet, most empirical studies infer homophily from proxies (e.g., co-following sources or group membership) without directly quantifying how close connected individuals are in their actual stances on ideological issues. 
This gap is consequential: societal divisions are fueled not just by who interacts with whom, but by how aligned those individuals' worldviews are. If online echo chambers are ultimately about shared ideology rather than just shared labels alone, a key step is to test whether ties are systematically associated with opinion similarity beyond what would arise from random interactions. In other words, do people preferentially engage with others holding similar opinions, and avoid those with divergent views, even within an online space?

Decades of research in psychology and communication suggest the answer is yes: individuals exhibit strong selective exposure and confirmation bias, leaning toward information and people that affirm their pre-existing beliefs \cite{stroud2010polarization,garrett2009echo}. Encountering conflicting viewpoints often induces cognitive dissonance \cite{festinger1962theory}, motivating people to ignore, dismiss, or counter-argue discordant messages rather than openly consider them \cite{lord1979biased,kunda1990case}. Consequently, on social media, users tend to curate their feeds (by choosing who to follow, friend, or engage with) in ways that filter out dissonant content \cite{bakshy2015exposure,mosleh2021cognitive}. This behavior can be understood as each individual having a ``latitude of acceptance'' around their current attitude \cite{sherif1961social}: messages or opinions falling within this latitude are deemed reasonable and receive attention, whereas those outside it are rejected or avoided. Likewise, the classic similarity-attraction effect \cite{byrne1971behavioral} affirms that people are interpersonally drawn to others with shared attitudes. This tendency is confirmed also in political contexts; for example, partisans show clear preference for like-minded partners in dating \cite{huber2017political}, and a recent field experiment found that shared partisanship dramatically increases the likelihood of forming a new social media connection \cite{mosleh2021shared}. These micro-level biases (seeking consonance and avoiding dissonance) provide a cognitive foundation for the persistence of homophilous clusters and echo chambers in networks.

Traditional quantitative value homophily models tend to associate a smooth decline of interaction frequency with the opinion distance \cite{axelrod1997dissemination,hohmann2023quantifying,bernardo2024asonam,lorenz2021individual,gestefeld2023calibrating}. 
A potential alternative to this distance-dependent influence decline model is to opt instead for a threshold-like discrimination between 
%acceptance and rejection of influence. 
interactions within and beyond a finite interaction horizon.
This approach is motivated by the fact that distance-dependent decay is often empirically noisy and heterogeneous, leading to ambiguities in the model-based quantifications. 
Our threshold-based approach is inspired by the field of opinion dynamics, where several formal agent-based models are equipped with explicit rules for threshold-based influence representation. In particular, the Bounded Confidence (BC) model \cite{hegselmann5others,deffuant2000mixing,bernardo2024bounded} posits a structural constraint where individuals only interact with (and influence) each other if their opinions differ by less than a certain threshold (which, in generalized formulations, can be heterogeneous or asymmetric); beyond that threshold, interaction ceases.
This simple rule, a stylized formalization of selective exposure, is normally used in a dynamical context, in which the influence of the neighbors is used to update the opinions. In this work we are not interested in dynamics of opinions, 
%but only in the neighbor selection mechanism that the BC concept suggests.
{but in the structural patterns of observed interaction neighborhoods in opinion space that the BC concept helps interpret.}

For this purpose we leverage recent social media data to 
%test whether interaction patterns in online ideologically-driven discourse are consistent with the BC selection rule.
{examine whether interaction neighborhoods in online ideologically-driven discourse display bounded and concentrated patterns in opinion space under a bounded confidence interpretation.}
Specifically, we analyze three heterogeneous datasets from Reddit and Twitter, covering different topical domains and interaction types, to examine whether users' interaction patterns reflect opinion-constrained engagement.

Each dataset provides two key ingredients: a reliable quantification of users' opinions, and an explicit network of interactions among those users. To avoid trivial confirmations of homophily that may arise from narrowly centered opinion distributions, we focus on datasets covering inherently adversarial or polarizing topics, where the opinions exhibit at least some deviation from unimodality, and in some cases a clear polarization. The three datasets considered are: (i)~Reddit-politics, a collection of Reddit user replies in a US politics subreddit, where user opinions are derived from sentiment analysis and replies to posts define directed interactions; (ii)~Twitter-Covid, a set of Twitter conversations on COVID-19, with tweet-level sentiments aggregated into user opinions, and replies as interaction edges; and (iii)~Twitter-contentious, a Twitter follower network centered on highly contentious political topics (e.g., abortion, gun control), where fact-checked hyperlinks approximate user ideologies and follower links indicate directed ties.
The functioning of these two platforms differs, as do the value and interpretation that can be given to both ``opinions'' and ``interactions''. 
Across these datasets, ``opinions'' are inferred by monitoring the user's activity on social media through different methods, like averaging the sentiment analysis on the collections of posts of a user (datasets i and ii), or fact-checking the content of the messages (dataset iii) (see \nameref{subsec:Data} for details). For what concerns ``interactions'', different regimes of tie strength are explored, from relatively weak ties (replies to a post, indicating context-dependent, transient interactions; datasets i and ii) to relatively strong ties (following a user, indicating identity-driven, durable, persistent interactions; dataset iii), allowing comparison of homophily also across tie strengths.
Detailed characterizations of the platform-specific interaction norms are provided in \textit{SI} Section~S6.

\begin{figure}[htbp]
    \centering
    \includegraphics[width=\linewidth]{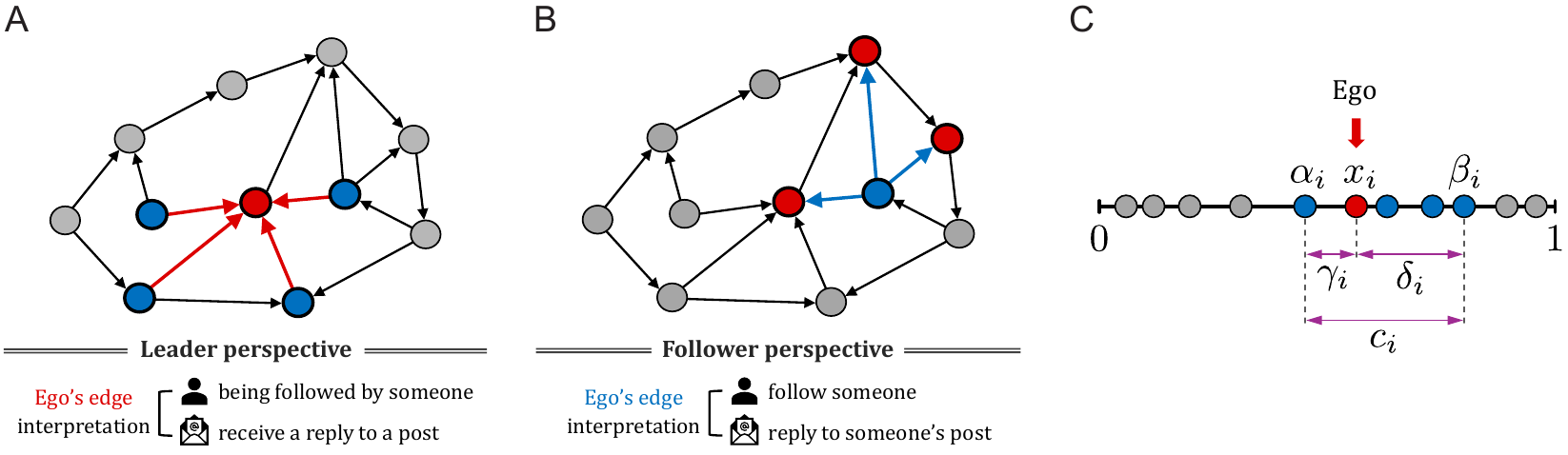}
    \caption{\textbf{Illustration of neighbor sets and indices.} (A):~Leader perspective. Ego as a leader (red node) with in-neighbors (blue nodes) representing followers or repliers. (B):~Follower perspective. Ego as a follower (blue node) with out-neighbors (red nodes) representing users that Ego follows or replies to. (C):~Schematic definition of the indices used to compute BC (see~\nameref{subsec:HomoCondition}). Ego's opinion is in red, while the opinions of its neighbors are in blue. \textit{Note:} the analysis in the main text adopts the follower perspective shown in Panel B, while the leader perspective shown in Panel A is investigated in the \textit{SI}.}
    \label{fig:sketches}
\end{figure}

Alongside comparing reply-based (weaker) and follow-based (stronger) networks of ties, we distinguish two perspectives when identifying a user's neighbors: a leader perspective (in which neighbors are in-neighbors) versus a follower perspective (with out-neighbors as neighbors), as illustrated in Fig.~\ref{fig:sketches}.
In the leader perspective, an individual (``Ego'') is characterized by who follows or replies to him/her, thus capturing ``Who engages with Ego?''.
In the follower perspective, the individual's neighbors are the users he/she follows or replies to, highlighting ``Who does Ego choose to engage with?''.
These viewpoints are complementary and reflect different directions of information flow and potential influence.
For the sake of readability, in the main paper we focus only on the follower perspective (Fig.~\ref{fig:sketches}B), 
%as it directly maps to the selection rule of the BC model, 
{as it aligns most naturally with our structural interpretation of bounded interaction neighborhoods,}
while in the \textit{SI} we present the corresponding results for the leader perspective (Fig.~\ref{fig:sketches}A).
{Since our analysis is based on unweighted user-user graphs, it characterizes who interacts with whom at least once, rather than how frequently such interactions occur.}
{Accordingly, our BC-inspired analysis does not specify any within-range interaction profile, but focuses on whether observed interaction neighborhoods exhibit bounded and concentrated structure in opinion space.}

By contrasting empirical interaction networks with degree-preserving null models, we address three core questions.
First, do users indeed engage with neighbors whose opinions more closely match their own than would occur under random reshuffling?
%, as postulated by the structural assumption of the BC model?
Second, if such opinion clustering exists, does it differ systematically between weaker reply-based ties and stronger follow-based ties?
Third, can additional factors, such as asymmetric {confidence ranges} (as suggested by the studies \cite{brandt2014ideological,zimmerman2024attraction}), characterize the directionality of opinion tolerance and capture more nuanced real-world behavior?
To answer these questions, we define several {interval-based structural} metrics (opinion gap, confidence range, mean/max opinion deviation, and range inclusion; see~\nameref{subsec:HomoCondition}) and evaluate how frequently users satisfy these conditions at both individual and population scales.
The effect of opinion similarity is isolated from the baseline network structure by benchmarking all metrics against degree-preserving (and, when relevant, confidence-range-preserving) randomized networks (see \nameref{subsec:DesignNull}).

%Our results confirm that the main structural features associated with BC are indeed pervasive in online social media interactions,
Our results show that {bounded and asymmetric patterns of homophily} are pervasive in online social media interactions, 
but highlight variability linked to tie strength and interaction type.
Weaker ties such as reply-based interactions generally exhibit moderate homophily with occasional cross-ideological exchanges, whereas stronger ties such as follow-based interactions substantially intensify like-minded opinion clustering and reinforce echo chambers.
%Additionally, we observe that individual tolerance ranges are often asymmetric, reflecting greater acceptance of moderate, mainstream-aligned opinions over more radical deviations from their current ideological stance.
%Additionally, we observe that individual tolerance ranges are often asymmetric, reflecting greater {engagement with} moderate, mainstream-aligned opinions {than with} more radical deviations from {the user's current stance}.
Additionally, we observe that individual tolerance ranges are often asymmetric, {with the direction of the asymmetry depending on the underlying opinion structure.}
%These findings provide evidence consistent with the structural characteristics of the BC framework in real-world online interactions and help characterize topological constraints associated with polarization and echo-chamber formation.
These findings provide {a structural characterization of online value homophily} and help clarify the topological constraints associated with polarization and echo-chamber formation.

\section*{Results}
\label{sec:Results}

The three datasets are hereafter denoted Reddit-politics, Twitter-Covid, and Twitter-contentious (see details in \nameref{subsec:Data} and \textit{SI}). Notice that to simplify the presentation, for most of this Results section we consider only one of the 24 months of Reddit-politics, and one of the 6 datasets of Twitter-contentious. The results are very similar for the remaining datasets, see \textit{SI}.
Furthermore, we focus only on the follower-perspective, i.e., on the interactions outgoing from a user, see Fig.~\ref{fig:sketches}B. The leader-perspective analysis yields similar results and is presented in the \textit{SI}.

\subsubsection*{Analysis at distribution level}
\label{subsec:Analysis_DistributionLevel}

\begin{figure}[htb!]
	\centering
	\includegraphics[width=0.9\linewidth]{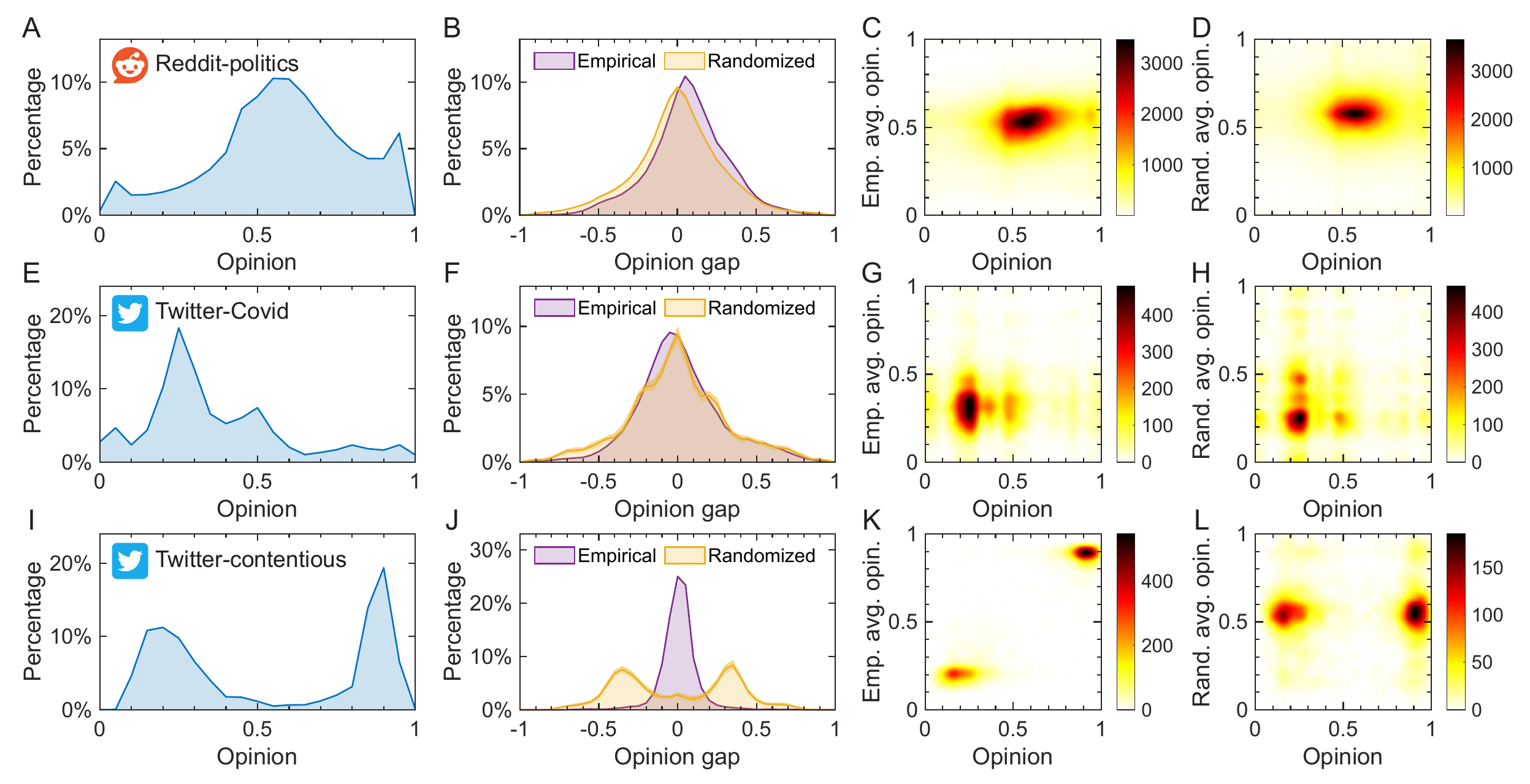}
	\caption{\textbf{Opinion distributions and {average-neighbor opinion comparison} across three datasets.}  Top row: Reddit-politics data. Middle row: Twitter-Covid data. Bottom row: Twitter-contentious data. Column 1: Opinion distributions. Column 2: Opinion gap distributions for the empirical network and the randomized null model. Column 3: Joint distributions of individual opinions and the average opinions of their empirical neighbors. Darker shading indicates higher density. Column 4: Corresponding joint distributions computed with randomized neighbor connections. 
	Randomized results are based on 20 independent trials. 
%	Column 5: Distributions of distance-dependent (``DD'') cases (i.e., Ego's neighbor count strictly decreases with opinion distance; see~\nameref{subsec:HomoCondition}) and non-distance-dependent (``NDD'') cases.
}
	\label{fig: case_2_opin_distribution}
\end{figure}

The empirical opinion distributions for the three datasets are shown in the left column of Fig.~\ref{fig: case_2_opin_distribution}. 
All datasets exhibit broad opinion patterns, with the Twitter-contentious dataset displaying the strongest bimodality.

To test whether individuals preferentially interact with neighbors holding similar opinions, we compare each individual's opinion with the average opinion of its first-order neighbors in the empirical social interaction graph.
For the three datasets, the resulting opinion gap $g$ (Eq.~\eqref{eq:opin-gap} in~\nameref{sec:Methods}) is shown in the second column of Fig.~\ref{fig: case_2_opin_distribution}. All three histograms are peaked around 0, meaning that Ego's opinions tend to be close to the average opinion of its neighbors, regardless of the overall opinion distribution.
This is particularly significant for the Twitter-contentious dataset, with its bimodal opinion distribution. 
The results are confirmed by the scatter plots shown in the third and fourth columns in Fig.~\ref{fig: case_2_opin_distribution}: in the Twitter-contentious dataset the bimodality pattern extends to the neighbor selection, clearly hinting at a high level of opinion polarization in the community.

The peaks of the opinion gaps $ g$ at 0 confirm the presence of homophily {\em on average} across the data.
%To investigate whether this homophily can be captured by commonly used principles, such as distance dependence, we test 1) how often the frequency of interactions decays in a monotone way with opinion distance, and 2) the correlation between distance among opinions and frequency of interaction. 
%The first measure is binary, and captures the cases in which strict monotonicity is observed, while the second is more relaxed and just describes the general trend. 
%As shown in the rightmost column of Fig.~\ref{fig: case_2_opin_distribution}, strictly monotonic decay is observed in only a small fraction of cases (3.0\% for Reddit-politics, 5.3\% for Twitter-Covid, and 8.7\% for Twitter-contentious).
%As shown in the \textit{SI} (Fig.~S1), the correlation between distance and interaction frequency is negative in about 75--87\% of the cases. Yet a closer inspection reveals that interaction patterns are normally very irregular and heterogeneous, containing wide fluctuations that leave little room for modeling the distance/frequency decay curves with simple smooth functions.
%See the lower panels in Fig.~S1 in the \textit{SI} for a few examples.
{However, average alignment alone does not characterize how neighbors are distributed in opinion space, and supplementary distance-decay diagnostics, reported in \textit{SI} Section~S4, remain heterogeneous across datasets and do not by themselves provide a stable basis for characterizing local neighborhood structure.}
This heterogeneity raises the question of what could be a reasonable way to quantify homophily in a manner that is both simple enough and interpretable under noisy, non-monotonic interaction profiles. 
The solution we consider is the BC-inspired interval summary described in Fig.~\ref{fig:sketches}C and in the \nameref{sec:Methods}: instead of fitting a specific distance-frequency function, we associate to each individual an interval containing all its neighbors and use it as a coarse descriptor of a bounded interaction horizon, without assuming any specific within-interval profile or functional form.
A necessary condition for this perspective to be meaningful is that these empirical intervals are narrower than expected by null models, which we verify next.

\begin{figure}[htb!]
	\centering
	\includegraphics[width=\linewidth]{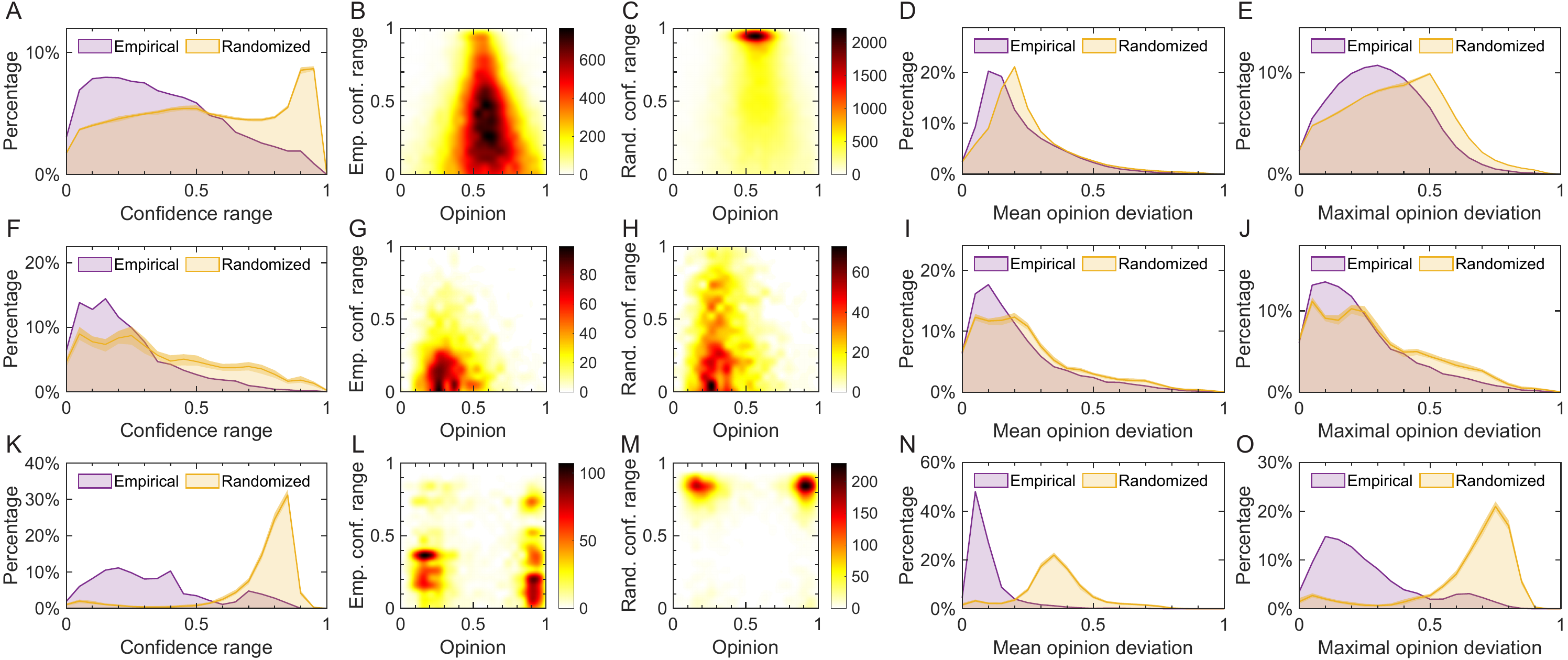}
	\caption{\textbf{Distributions of confidence range, mean deviation, and maximal deviation across three datasets.}  Top row: Reddit-politics data.  Middle row: Twitter-Covid data. Bottom row: Twitter-contentious data. Column 1: Confidence range distributions for the empirical network and the randomized null model. Column 2: Joint distributions of individual opinions and their empirical confidence ranges. Darker shading indicates higher density. Column 3: Corresponding joint distributions using randomized confidence ranges. Column 4: Mean opinion deviation distributions. Column 5: Maximal opinion deviation distributions.
	 Randomized results are based on 20 independent trials.}
	\label{fig: case_2_conf_range_distribution}
\end{figure}

If instead of looking at the average of the neighbors' opinions we look at their dispersion, then we can consider a measure such as the confidence range $c$ (Eq.~\eqref{equ: Ci}). Figure~\ref{fig: case_2_conf_range_distribution} shows that in all three datasets the confidence range distributions are skewed towards 0, meaning that the neighbors' opinions tend to be concentrated rather than scattered along the entire opinion axis. In all three cases the null models tend to produce larger ranges, and the difference is statistically significant ($p<0.05$), with moderate effect sizes in Reddit-politics (Cohen's effect size $d \approx 0.63$) and Twitter-Covid ($d \approx 0.59$), see null model distributions on the left column and scatter plots in the second and third columns of Fig.~\ref{fig: case_2_conf_range_distribution}. 
{For the representative datasets shown in the main text, the empirical confidence ranges are also quantitatively small at the dataset level; see Table~S2 in the \textit{SI} for the corresponding median and interquartile range values across datasets and perspectives.}

%Having a narrow confidence range is indicative of a neighbor selection process which is focused rather than scattered, but by itself it is not sufficient to express a preference for a BC-style thresholded selection over a generic distance-based homophily.
Having a narrow confidence range is indicative of {an interaction neighborhood} that is focused rather than scattered, but by itself it is not sufficient to {characterize the local structure of interaction neighborhoods.}
What is required in addition is that the Ego's opinion $ x_i $ falls \textit{within} the range of the neighbors' opinions $ [\alpha_i, \beta_i]$. 
To check how users conform to this more strict principle, we examine several  opinion-based metrics: mean opinion deviation (Eq.~\eqref{equ:MeanDeviation}), maximal opinion deviation (Eq.~\eqref{equ:MaxDeviation}), and, later on, a range inclusion binary condition (Eq.~\eqref{equ:RangeInclusion}).

The two rightmost columns of Fig.~\ref{fig: case_2_conf_range_distribution} show that across all datasets, users tend to have significantly smaller mean and maximal opinion deviations from their neighbors compared to the randomized cases ($p<0.05$).
To quantify the strength of this restriction, we calculate Cohen's $d$ effect sizes.
In Reddit-politics, the effect sizes for mean ($\Delta_{\text{mean}}$) and maximal ($\Delta_{\text{max}}$) deviations are $d \approx 0.23$ and $d \approx 0.34$, respectively, while in Twitter-Covid, they are $d \approx 0.22$ and $d \approx 0.27$.
Despite the moderate magnitude, these values indicate a consistent tightening of the opinion neighborhood relative to random expectation that is statistically robust across large datasets.
These trends are especially pronounced in cases where strong ties align with high polarization (i.e., the Twitter-contentious data).
Similar results are obtained in other months of the Reddit-politics dataset and for the other topics of the Twitter-contentious dataset, see Figs.~S4 and S5 in the \textit{SI}.
Comparative analyses from the leader perspective are also included in the \textit{SI} and show consistent trends, see Figs.~S8 and S9 in the \textit{SI}.
Together, these observations suggest that real-world interaction neighborhoods are, on average, more concentrated in opinion space than expected under the null baselines, even in reply-based networks, with this effect being most pronounced in follow-based networks.

Summarizing, the distribution-level evidence 
%is consistent with an effective BC-style structural rule for neighbor selection: 
indicates that interaction neighborhoods are typically narrower (in range) and more centered (in deviations) than expected under the null baselines.

%\subsubsection*{BC satisfaction rates at population-level}
\subsubsection*{{Population-level rates of interval-based conditions}}
\label{subsec:Population-levelAnalysis}

%We now examine BC satisfaction rates by introducing three population-wide rates $R_1$, $R_2$, and $R_3$ (Eqs.~\eqref{equ:MeanDeviationRate}--\eqref{equ:RangeInclusionRate}). $ R_1 $ and $ R_2 $ quantify the extent to which individuals satisfy mean and max opinion deviation conditions compared to null expectations.
We now examine three population-wide rates $R_1$, $R_2$, and $R_3$ (Eqs.~\eqref{equ:MeanDeviationRate}--\eqref{equ:RangeInclusionRate}), which quantify the extent to which {the corresponding interval-based conditions} are satisfied at the population level relative to null expectations.
Note that the randomized interactions here are generated using a refined range-based null model (see \nameref{subsec:DesignNull}), which preserves 
% each individual's empirical confidence range 
{the scale of the empirical interaction horizon} and in this way provides a fairer baseline than the plain randomized one.
Here, $R_1$ and $R_2$ measure population-level rates associated with mean and maximal opinion deviation, respectively, while $R_3$ is a population-level indicator of range inclusion, i.e., whether the opinion of a user is contained within the observed neighbor interval formed by its neighbors; see Eq.~\eqref{equ:RangeInclusionRate}.

\begin{figure}[htb!]
	\centering
	\includegraphics[width=0.95\linewidth]{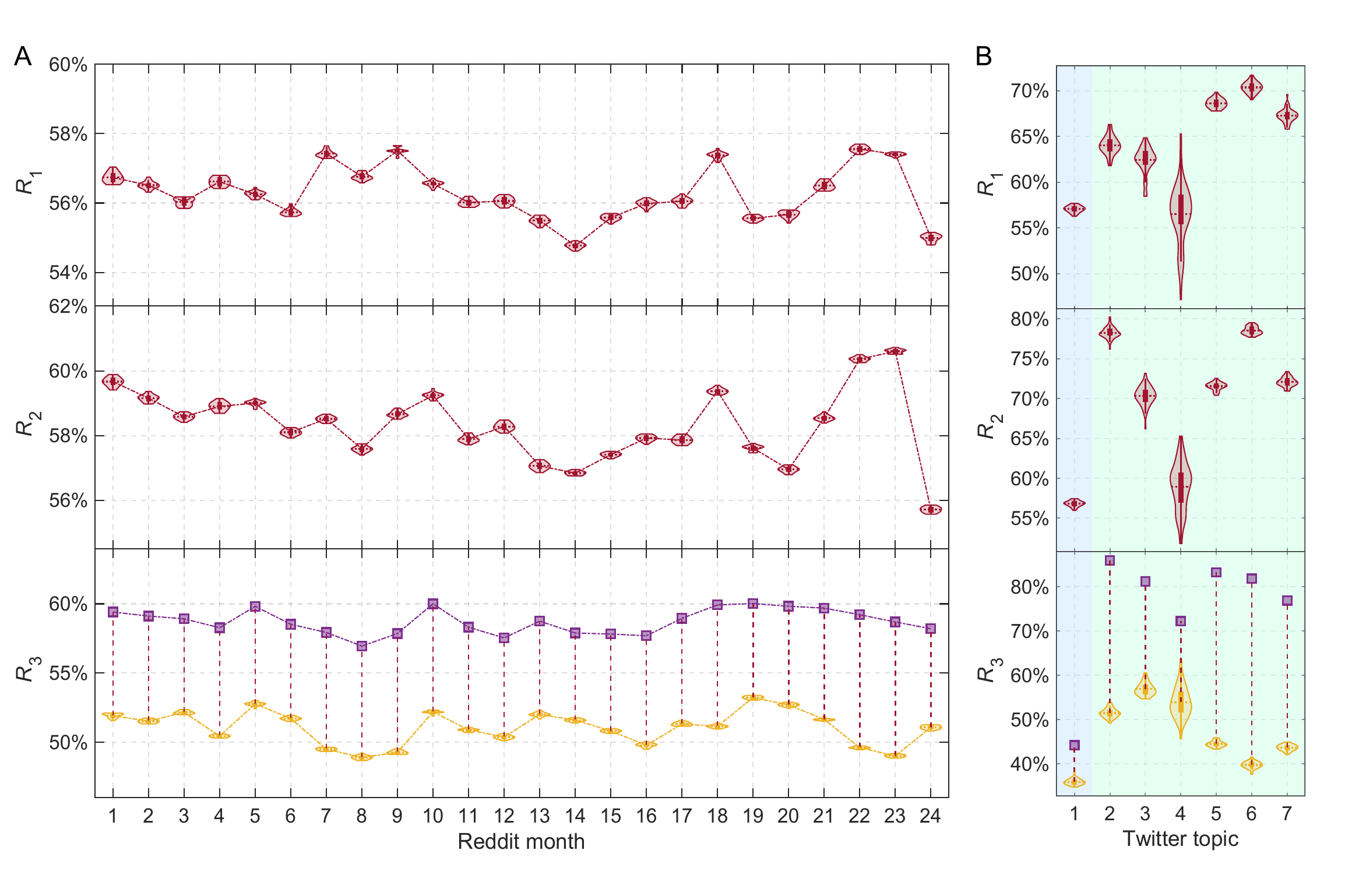}
	\caption{\textbf{{Population-level rates of interval-based conditions} in Reddit and Twitter datasets.}
		(A):~Longitudinal trends for $R_1$, $R_2$, and $R_3$ over 24 months in the Reddit-politics data.
		(B):~Comparisons of these rates across multiple Twitter topics. 
		The first dataset (labeled ``1'', Twitter-Covid) is reply-based, while the remaining six datasets (labeled ``2''--``7'', Twitter-contentious topics) are follow-based. The violin plots in the first two rows show how $ R_1 $ and $ R_2 $ evolve over 20 random instances of the range-based null model, while the last row compares the empirical and randomized $ R_3 $ (the latter, in yellow, over repeated instances represented as violin plots).}
	\label{fig: case_2_Reddit_Population-level_satisfaction_rates_new}
\end{figure}

The three indicators $R_1$, $R_2$, and $R_3$ can be compactly expressed at the population level, hence in Fig.~\ref{fig: case_2_Reddit_Population-level_satisfaction_rates_new} we can show them for all 24 months of the Reddit-politics and 6 topics of the Twitter-contentious dataset. 

In Reddit-politics, for all 24 months, $R_1$ is around 55--58\%, $R_2$ around 56--60\%, and $R_3$ around 57--60\%, indicating moderate but persistent homophily.
Twitter-Covid presents similar $R_1$ and $R_2$ but a lower $R_3$ ($<50\%$), implying limited range inclusion, yet still surpassing the random baselines.
Across all metrics, these results quantify the prevalence of {homophilic structure of interaction neighborhoods} beyond baseline random expectations. Specifically, $R_1$ and $R_2$ consistently exceed the 50\% benchmark, indicating neighborhoods strictly tighter than random, while the empirical $R_3$ substantially surpasses the explicit randomized baseline (see the gap between purple and yellow values in Fig.~\ref{fig: case_2_Reddit_Population-level_satisfaction_rates_new}).
By comparison, Twitter-contentious exhibits even higher values across all three indices, generally with $R_1>60\%$, $R_2>70\%$, and $R_3>80\%$, reflecting strong opinion similarity among interacting individuals also at the population level.
The main reason for the lower $R_3 $ rate in the Twitter-Covid data is that the distribution of empirical confidence ranges is significantly more peaked at low $c$ values than in the other datasets (see the left column of Fig.~\ref{fig: case_2_conf_range_distribution}), making it more likely to fail the range inclusion test (a binary test).
In fact, as can be observed in Fig.~\ref{fig: case_2_conf_range_distribution}, both mean and max opinion deviations are similar to the other datasets, namely they are highly skewed towards 0. 
Similar findings under the leader perspective are presented in Fig.~S10 of the \textit{SI}.

%\subsubsection*{Variation of BC satisfaction rates across opinion intervals}
\subsubsection*{{Variation of interval-based rates across opinion intervals}}

\begin{figure}[htb!]
	\centering
	\includegraphics[width=\linewidth]{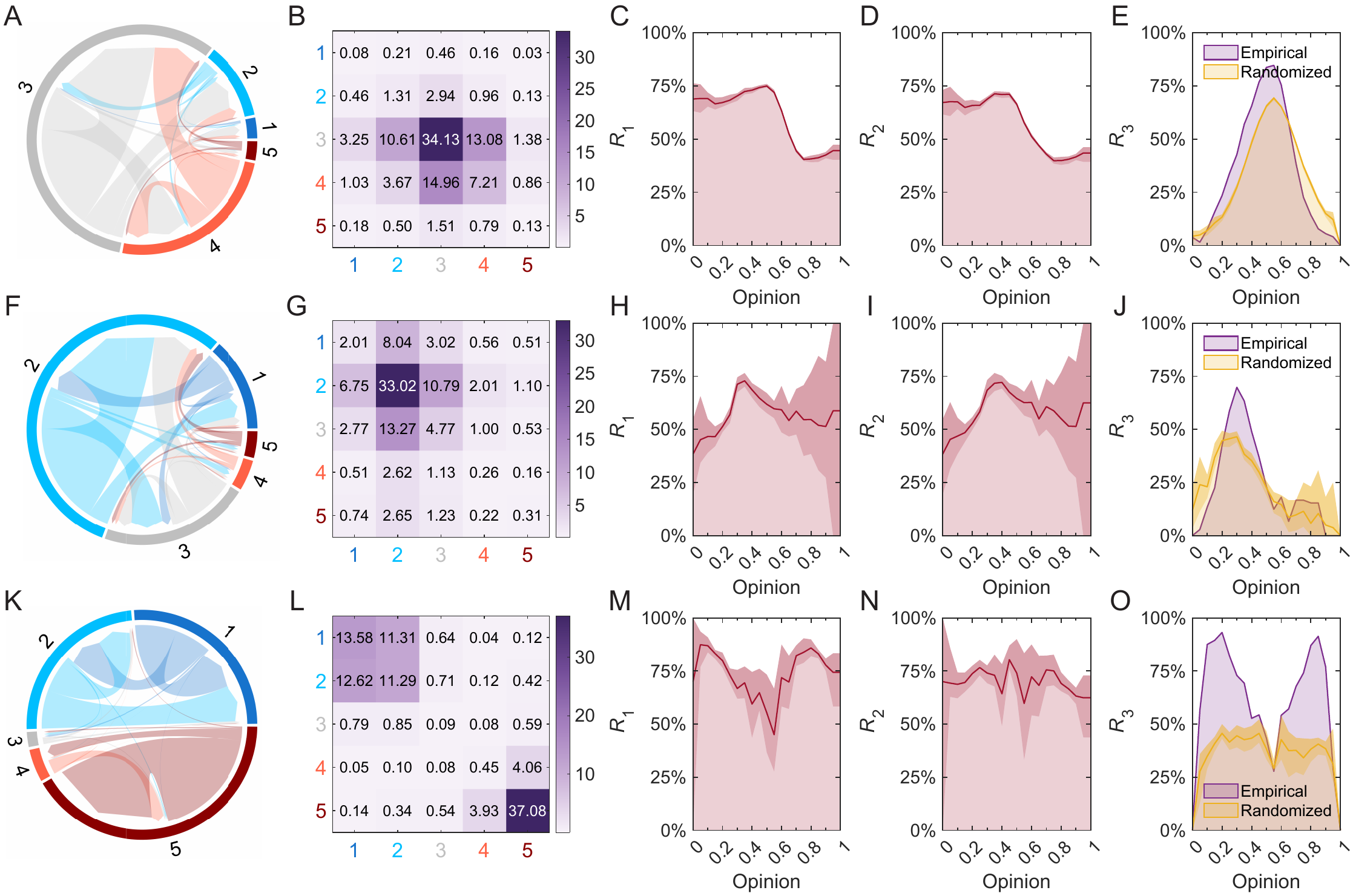}
    \caption{\textbf{Interaction patterns and {rates of interval-based conditions} across partition-based opinion intervals in three datasets.}
    	Top row: Reddit-politics data. Middle row: Twitter-Covid data. Bottom row: Twitter-contentious data.
    	Column 1: Chord diagrams illustrating the connectivity both within and between the groups corresponding to five ordered opinion intervals: ``1'' [0,0.2], ``2'' (0.2,0.4], ``3'' (0.4,0.6], ``4'' (0.6,0.8], and ``5'' (0.8,1]. The angular size of each sector reflects the total interaction volume associated with that opinion interval (see also \textit{SI} Table~S1 and Fig.~S3 for a more detailed analysis), while chord thickness indicates the volume of cross-interval interactions.
    	Column 2: Heatmaps representing the interaction percentages between opinion intervals. The $(i,j)$-th cell denotes the proportion of interactions originating from opinion interval $i$ and directed to opinion interval $j$, normalized by the total number of interactions. Darker shades indicate higher interaction frequencies.
    	Columns 3--5: {Rates for} mean deviation $R_1$, maximal deviation $R_2$, and range inclusion $R_3$ over 20 range-based randomized trials.}
	\label{fig: case_2_Individual-level_satisfaction_rates_new}
\end{figure}

In order to examine how {the interval-based rates} vary across different regions in opinion space, we partition users into distinct opinion intervals.
Figure~\ref{fig: case_2_Individual-level_satisfaction_rates_new} shows that in Reddit-politics and Twitter-Covid, these rates peak around the dominant local center in opinion space (approximately $x_{i} \approx 0.55$ and $x_{i} \approx 0.3$, respectively).
Note that even in these datasets with unimodal opinion distributions, substantial {values of the interval-based rates} are observed across most opinion intervals when assessed relative to the range-based null baselines. 
%This indicates that the preference for bounded confidence interaction is a robust local feature, 
This indicates that {bounded and non-random local concentration} is a robust feature across most opinion intervals,
distinguishable from the baseline geometry.
In contrast, in Twitter-contentious, which exhibits a bimodal distribution, these rates are generally highest at both extremes of the opinion distribution.

The chord diagrams and heatmaps in Fig.~\ref{fig: case_2_Individual-level_satisfaction_rates_new} illustrate further dataset-level differences in interaction patterns. 
The first thing to notice is that in Twitter-Covid the left-leaning opinion intervals (``against vaccination'') generate the largest volume of interactions, and these seldom cross the opinion divide.
%Furthermore, in Reddit-politics and to some extent also in Twitter-Covid,  moderate or central opinion intervals tend to bridge users from different ideological segments, resulting in more frequent cross-interval engagement.
Furthermore, in Reddit-politics and to some extent also in Twitter-Covid, more central opinion intervals tend to bridge users from different opinion segments, resulting in more frequent cross-interval engagement.
These observations still hold even after the effect of the opinion distribution has been subtracted; see Fig.~S2 in the \textit{SI}.
By contrast, in Twitter-contentious, interactions remain largely constrained within extreme opinion segments, reinforcing segregation. 
% in Twitter-contentious, interactions remain largely constrained within extreme opinion segments, reinforcing segregation.
This suggests that while reply-based interactions (as in Reddit-politics and Twitter-Covid) may allow somewhat greater exposure to diverse opinions, stronger follow-based relationships (as in Twitter-contentious) may lead to more entrenched opinion groups and echo chambers.
Additional supporting results across multiple Reddit months and other Twitter topics are provided in Fig.~S6 of the \textit{SI}, with corresponding leader-perspective comparisons shown in Fig.~S11 of the \textit{SI}.

\subsubsection*{{Asymmetric neighborhood spans}}
\label{subsec:Asymmetric}

The previous analysis reveals that 
%BC satisfaction rates 
{the rates of the interval-based conditions} vary across different opinion regions.
Furthermore, the {observed neighbor intervals} $ [\alpha_i, \beta_i] $, even when they contain the opinion $ x_i $, are rarely centered around $ x_i$. 
Both features call for a more fine-grained examination of the 
%tolerance thresholds of the individuals. 
{directional structure of local interaction neighborhoods.}
%For that we consider the left and right confidence ranges $ \gamma_i $ and $ \delta_i $ (Eq.~\eqref{eq:gamma-delta}) and the asymmetry index $ s_i $ defined in Eq.~\eqref{equ:AsymmetryIndex}. 
For that we consider the left and right {offsets $\gamma_i$ and $\delta_i$ of the observed neighbor interval relative to Ego} (Eq.~\eqref{eq:gamma-delta}), together with the asymmetry index $s_i$ defined in Eq.~\eqref{equ:AsymmetryIndex}.

\begin{figure}[htb!]
	\centering
	\includegraphics[width=\linewidth]{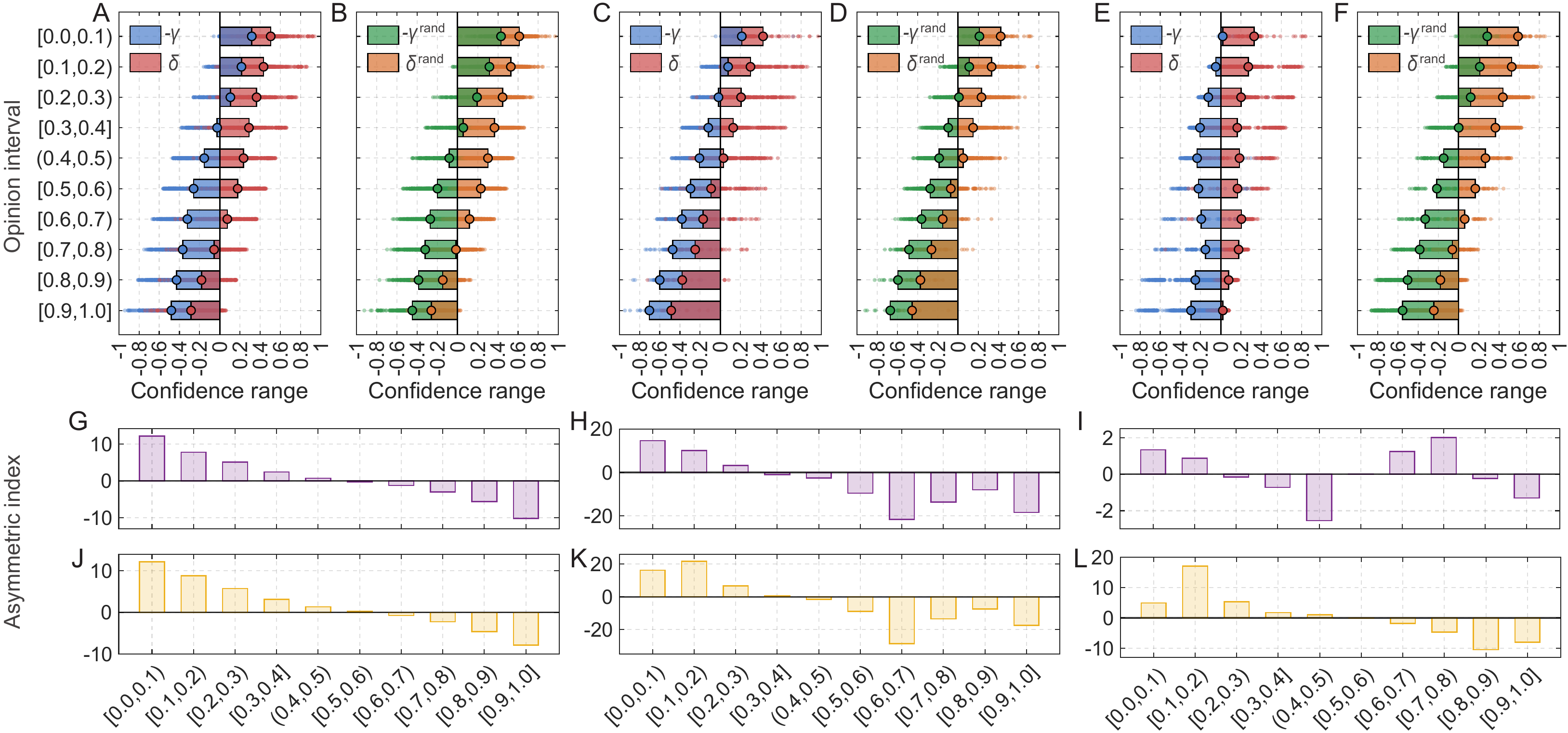}
	\caption{\textbf{Asymmetric {neighborhood spans} and opinion-dependent tolerance patterns across three datasets.}
		(A--F) Distributions of negated left {offsets} ($-\gamma_i$) and right {offsets} ($\delta_i$) for different opinion intervals in empirical data (A, C, E) and their corresponding range-based null models (B, D, F) over 20 randomized trials.
		Circular markers denote the mean values within each interval.
		(G--I) Asymmetry index $s_i$ across different opinion intervals in empirical data.
        (J--L) Corresponding asymmetry index $s_i$ from range-based null models.}
	\label{fig: case_2_fig_add_LeftRightC_new}
\end{figure}

Figure~\ref{fig: case_2_fig_add_LeftRightC_new} shows the asymmetry of the left/right {signed offsets} across opinion intervals, revealing distinct patterns across datasets.
In Reddit-politics and Twitter-Covid, individuals with extreme opinions exhibit strong {directional} asymmetry, 
%with larger confidence ranges toward the ideological center and reduced ranges toward further extremism in their own direction.
{with the observed neighbor span often extending further toward positions that are locally more mainstream in the corresponding dataset.}
%This pattern aligns with higher BC satisfaction rates at $x_{i} \approx 0.5$ in Reddit-politics and $x_{i} \approx 0.3$ in Twitter-Covid, reflecting their respective ideological centers.
This pattern aligns with higher {interval-based concentration around} $x_i \approx 0.5$ in Reddit-politics and $x_i \approx 0.3$ in Twitter-Covid, reflecting their respective ideological centers. 
By contrast, Twitter-contentious, characterized by a bimodal opinion distribution, exhibits weaker {and more segment-dependent} asymmetry.
Individuals near ideological attractors ($x_i \approx 0.2$ and $x_i \approx 0.9$) show comparable {local spans} within their respective segments.
In Twitter-contentious, the asymmetry index $s_i$ remains small {in magnitude} across all opinion intervals, whereas Reddit-politics and Twitter-Covid display large $|s_i|$ values at extreme positions, 
%reinforcing the idea of a strong directional bias in tolerance.
indicating stronger directional bias in {the observed neighborhood span.}

This distinction becomes more evident when compared to the null model. 
In Reddit-politics and Twitter-Covid, empirical and randomized cases exhibit similar broad patterns, indicating that opinion distributions 
%to a large extent shape individual asymmetry.
{and geometric constraints already account for a large part of the observed asymmetry.}
%However, a clearer residual pattern emerges and it is shown in \textit{SI} Section S4 (Fig.~S2): subtracting the null baseline reveals
{A complementary residual view at the interaction-pattern level is provided by Fig.~S2 in the \textit{SI}, which shows} non-random structure beyond what is expected from the marginal opinion distribution alone.
In contrast, in Twitter-contentious, empirical results deviate significantly from the null model, suggesting that social network structure, particularly long-lasting follow relationships, is associated with stronger within-segment 
%asymmetry while sharpening separation from opposing opinions.
{directional organization and clearer separation from opposing segments.}
Similar effects are seen across other months of the Reddit-politics dataset and Twitter topics (see Fig.~S7 in the \textit{SI}), with leader-perspective comparisons shown in Fig.~S12 in the \textit{SI}.

\section*{Discussion}
\label{sec:discussion}

%Our study offers empirical evidence for BC neighbor selection mechanisms in large-scale online social networks.
Our study provides large-scale empirical evidence for {bounded and non-random structural patterns of opinion homophily} in online social networks.
Despite variations in platform design, user demographics, and topical context, all three datasets consistently show that 
%individuals preferentially engage with others whose opinions deviate less from their own;
{interaction neighborhoods are more concentrated in opinion space} than expected under randomized baselines.
%that is, the observed opinion confidence range between connected users is significantly smaller than that of a randomized network, which is 
In particular, the observed confidence ranges and opinion deviations are systematically smaller than those of randomized networks,
a pattern consistent with the pervasive ideological homophily reported on social media \cite{halberstam2016homophily,barbera2015birds,conover2011political}.  
%This pattern aligns with selective exposure theory \cite{stroud2010polarization,garrett2009echo}, supporting the idea that active neighbor selection (as occurs in the follower-perspective investigated above) is a robust structural feature of social alignment.
This pattern also resonates with selective exposure theory \cite{stroud2010polarization,garrett2009echo}, and points to {actively formed interaction neighborhoods}, as captured by the follower perspective investigated above, as an important structural feature of social alignment.

%Conceptually, most value homophily representations are characterized by a continuous and monotonic decline in tie probability as opinion distance increases, turning homophily into a ``smooth'' distance-based effect.
Conceptually, many quantitative representations of value homophily are characterized by a continuous and monotonic decline in tie probability as opinion distance increases, turning homophily into a ``smooth'' distance-based effect \cite{hohmann2023quantifying,bernardo2024asonam,lorenz2021individual,gestefeld2023calibrating}.
{In the present work, however, we do not attempt to identify a unique interaction kernel or to infer a precise functional form of distance dependence from the data.}
%By contrast, BC entails a thresholded acceptance: interactions occur only within a (potentially asymmetric) tolerance window and are absent outside it.
{Instead, from a bounded confidence perspective, we focus on whether interaction neighborhoods exhibit two structural features in opinion space: a finite interaction horizon and non-random concentration within that horizon.}
%Clearly, for the BC model to be useful, these windows
Clearly, {for this perspective to be meaningful, these horizons} have to span only a limited portion of the opinion space.
%The BC model also implies a ``range inclusion'' property,
{A related structural feature is the} ``range inclusion'' property,
which requires an Ego's opinion to lie within the span of its neighbors’ opinions.
Empirically, our use of a range-preserving null model and the range inclusion index ($R_3$)
% directly test this windowed acceptance,
{quantify this local consistency,}
while the asymmetry index ($s_i$) and the contrast between follow-based and reply-based networks (discussed below) reveal directional and tie-dependent selectivity in the data.
%not predicted by distance-dependent  assortativity.

An important insight of our work is the contrast between reply-based (weaker tie) and follow-based (stronger tie) networks; that is, the strength of social ties clearly modulates ideological clustering. 
While both manifest homophily, the effect is substantially stronger in the follow-based interactions.
In these settings, users continually receive information from ideologically similar sources they have themselves chosen, reinforcing echo chambers.
This is in accord with experiments demonstrating that people are far more likely to connect with politically similar others than with opponents \cite{mosleh2021shared}, and with network studies suggesting that stable ties magnify polarization \cite{min2020underneath,ross2022echo}.
By comparison, reply-based networks include a nontrivial fraction of cross-ideological exchanges.
While some of these may be of confrontational nature \cite{de2021no,gaisbauer2021ideological}, they may still in part show exposure to different opinions and may temper extreme clustering \cite{bakshy2015exposure,garimella2017reducing}.
This discrepancy may even hint at a potential side effect of tie strength: whereas stronger ties can intensify polarization, weaker ties may enable some level of cross-ideological exposure that, despite potential negativity, provides at least some heterogeneity of thought.
Notably, the most polarized issues (our Twitter-contentious dataset) exhibit the strongest homophily patterns, suggesting that a bimodal opinion structure combined with strong in-group ties provides ideal conditions for 
%BC-like segregation.
{strongly segregated interaction neighborhoods.}

It is worth remarking that for the current datasets, the amount of antagonism is not quantified 
%by the sentiment analysis tools. 
{by the opinion inference tools.}
We expect this to be a confounding factor especially for the reply-based networks, which are likely to contain a certain fraction of confrontational content~\cite{de2021no,gaisbauer2021ideological}. For instance, interactions with opinions that are not closely aligned may be due to critical replies or challenges, rather than endorsements. 
This heterogeneity may explain why 
%interaction probability sometimes peaks at non-zero distances (as noted in \cite{gaisbauer2021ideological} and hinted at in our \textit{SI} Fig.~S1).
{the relation between opinion distance and neighbor count is sometimes irregular and may even peak away from zero (as noted in \cite{gaisbauer2021ideological} and illustrated in our \textit{SI} Section~S4).}
The issue is likely less relevant for follow-based networks, where users typically endorse users they are aligned with.

Nevertheless, the persistence of a bounded interaction horizon across datasets supports the robustness of a finite engagement range: users engage, whether to agree or to argue, primarily within a limited opinion span.
Thus, our results provide evidence of bounded engagement, a structural pattern 
%consistent with the assumptions of influence-based BC-type models.
{that aligns with the bounded-neighborhood intuition underlying BC-type models.}
In this respect, the potential presence of antagonism only reinforces the significance of our findings as it makes our arguments conservative: by mixing attitudinal homophily with adversarial exchanges, we are likely underestimating the strength of pure supportive alignment.

When instead of the follower-perspective investigated above we consider the leader-perspective, the results are qualitatively similar for basically all indicators, see Figs.~S8--S12 in the \textit{SI}.
In the follower perspective, users actively choose who to follow or to whom they reply, whereas in the leader perspective, popular accounts passively attract neighbors. Overall, these leader-perspective results reinforce the notion that being commented on can dilute homophily signals compared to actively choosing whom to engage, especially in contexts, like political threads, that attract both supportive and adversarial replies.
Consistent with this, prior research finds that active interaction produces more segregated clustering, while passive reception allows a broader mix of engagement, potentially including critics or random commenters \cite{bakshy2015exposure,knobloch2015selective,dvir2017media}. Although conceptually different, both perspectives yield consistent evidence for a bounded-engagement structure in the data: active self-selection and passive audience formation collectively characterize an interaction topology that is 
%significantly more ideologically homogeneous than expected from null models. 
significantly more homogeneous in opinion space than expected from null models.

In addition, the study reveals that tolerance for opinion divergence is often asymmetric. Many users are more open to opinions on one side of their own stance than the other, reflecting a biased personal latitude of acceptance \cite{sherif1961social}. 
This effect is particularly pronounced in datasets based on replies (Reddit-politics, Twitter-Covid), in which users at both 
% ideological extremes more readily interact with moderates than with radicals on their own side.
{extremes of the opinion distribution} more readily interact with {more central, locally mainstream positions} than with more radical positions on their own side.
Part of this bias arises mechanically from opinion distributions: extremists tend to have more centrist neighbors simply because there are more moderates available.
However, as shown in the \textit{SI} Fig.~S2, even after accounting for exposure in the null model, users tend to have more interactions toward center groups.
In some cases, these effects seem to contribute to the asymmetric patterns we detect: on Reddit, far-right users show an even stronger attraction to moderates than the null model predicts; 
in contrast, on Twitter-Covid certain anti- and pro-vaccine groups exhibit a weaker asymmetry than expected.
Interpreting these phenomena is difficult without detailed content analysis.
Meanwhile, in the highly polarized Twitter-contentious dataset, tolerance asymmetry is minimal, 
%and users are essentially anchored to one of the ideological attractors.
and users appear to remain close to one of the ideological attractors.
%The marked asymmetry observed supports generalized BC formulations \cite{hegselmann5others,douven2022network,hegselmann2023bounded,bernardo2024bounded}, challenging the common simplification of symmetric thresholds and confirming instead that real-world users operate with distinct, direction-sensitive confidence bounds driven by confirmation bias~\cite{nickerson1998confirmation} and motivated reasoning~\cite{kunda1990case}.
{The marked asymmetry observed is consistent with generalized BC formulations \cite{hegselmann5others,douven2022network,hegselmann2023bounded,bernardo2024bounded}, and suggests that directional tolerance in real data is often shaped by the underlying opinion structure. 
In particular, what counts as a locally mainstream position need not coincide with the neutral midpoint, especially in bimodal settings where local ideological attractors emerge.}

It is worth recalling that Reddit and Twitter allow significantly different interaction rules, reflected in the network topologies in the datasets.
Reddit-politics is essentially a large forum where all users participate in common threads, leading to broader exposure and more uniform degree distributions (see Fig.~S3 in the \textit{SI}).
Related work also suggests that Reddit's single-forum structure can foster more heterogeneous exposure \cite{de2021no}.
In contrast, Twitter's discourse is organized into personalized feeds, where degree distributions reflect the opinion distributions themselves, thereby reinforcing echo chambers.
On user-centric platforms like Twitter, tie formation may also reflect platform-specific norms and non-ideological factors (e.g., information seeking or strategic attention), which can introduce structural noise independent of opinion alignment.
In the topic-specific follow graphs analyzed here, we observe zero reciprocity (no mutual follower pairs); this is a property of the restricted graphs induced by the dataset construction \cite{hohmann2023quantifying} and should not be interpreted as a general statement about reciprocity on Twitter (see \textit{SI} Section~S6).
Nevertheless, the strong clustering observed relative to degree-preserving null baselines is consistent with cognitive alignment being an important contributor to the resulting topology, although we cannot fully disentangle it from platform- and topic-specific effects.
These observations suggest that the echo chamber effect can be amplified or mitigated by platform architecture.

The algorithmic recommendation system employed by platforms is clearly a confounding factor~\cite{santos2021link,chavalarias2024single}. We acknowledge that observational data cannot strictly disentangle user-level cognitive choices from platform constraints: a user might theoretically possess a wider cognitive tolerance but fail to connect with divergent opinions due to stochastic effects, scarcity of such opinions in the network, or algorithmic filtering. 
However, since recommender systems typically optimize engagement by learning and reinforcing pre-existing user preferences, algorithms may plausibly act as amplifiers of cognitive tendencies rather than independent creators of barriers. 
%Our findings thus characterize the ``effective'' bounded confidence, i.e., the realized interaction range resulting from the interplay between cognitive preference, adversarial motivation, and the underlying network topology.
Our findings thus characterize an ``effective'' bounded engagement range, i.e., the realized interaction span resulting from the interplay between cognitive preference, adversarial motivation, and the underlying network topology.

Other confounding factors potentially influencing our results include the possible presence of extra interactions (off-line or on other forums) and the temporal evolution of opinions.
Concerning the latter, it is worth remarking that while the Reddit dataset allows longitudinal analysis \cite{pansanella2022change}, in this work we focus specifically on the static structural features of bounded {interaction neighborhoods.}
The full (dynamical) BC model conceptually comprises two distinct steps: a neighbor selection rule based on confidence bounds, and an opinion update rule based on averaging.
Assessing the full dynamic prediction would therefore require tracing individual belief trajectories over time.
However, an ``opinion'' is a complex and multidimensional construct~\cite{mason2015disrespectfully,luan2025coevolutionary}, 
%and any ``opinion change'' arises from complex factors and cognitive processes not reducible to short-lived Reddit interactions, making the investigation of the update rule somewhat implausible in this context.
and observed opinion change may arise from multiple factors and cognitive processes beyond the interactions recorded in these datasets.
{In fact, a dynamical analysis of the longitudinal Reddit dataset was attempted in \cite{pansanella2022change}, but yielded only limited support for BC-style update dynamics in these data.}
For this reason, our results focus on characterizing the selective exposure {patterns} that we see in the data, without making claims about the opinion update rule or temporal causality.

Our findings also suggest strategies for mitigating polarization. It is known that unfiltered exposure to extreme out-group content can backfire \cite{bail2018exposure}, so interventions should emphasize more moderate cross-cutting encounters \cite{garimella2017reducing,matakos2017measuring}.
For example, platforms might introduce content positioned at the periphery but strictly within a user's current confidence interval, thereby leveraging the intuition from BC-type attraction dynamics to maximize potential attitude change without triggering rejection \cite{axelrod2021preventing,helberger2018exposure}.
Diversifying feeds in small steps could expand comfort zones without provoking backlash \cite{munson2010presenting}.
Another strategy is to foster cross-cutting ``weak tie'' interactions \cite{bakshy2015exposure}, leveraging common-ground discussions to bridge communities \cite{tan2016winning}.

%In conclusion, our research provides empirical evidence consistent with the bounded confidence principle in online social networks, demonstrating how tie strength, issue polarization, and asymmetric opinion tolerance jointly characterize homophily patterns.
In conclusion, our research shows that tie strength, issue polarization, and asymmetric opinion tolerance jointly shape {bounded and non-random patterns of online value homophily.}
By disentangling the roles of weaker vs.\ stronger ties, we underline the critical importance of persistent social connections in reinforcing echo chambers, while also highlighting opportunities for cross-ideological interactions in reply-based contexts.
Future research should extend this framework to other platforms and multi-issue opinion spaces~\cite{ojer2025charting}, while integrating finer-grained user attributes~\cite{brady2017emotion}.
Theoretical refinements also remain of interest, such as understanding {how empirically identified interaction ranges may relate to dynamic models of opinion formation}.
%estimating the precise functional form of distance dependence~\cite{gaisbauer2021ideological}.
Finally, investigating the interplay between user actions and algorithmic recommendations~\cite{sirbu2019algorithmic} may advance our understanding of online homophily and its consequences.

\section*{Methods}
\label{sec:Methods}

%\subsection*{BC-based neighbor selection conditions for opinion homophily}
\subsection*{{Interval-based structural measures} of opinion homophily}
\label{subsec:HomoCondition}

We consider a social network represented as a directed graph $\mathcal{G} = \{\mathcal{I}, \mathcal{E}\}$, where $\mathcal{I}=\{1,2,\ldots,N\}$ is the set of $N$ individuals, and $\mathcal{E} \subseteq \mathcal{I} \times \mathcal{I}$ is the set of directed interactions between them. 
A directed edge $(i, j)$ indicates that individual $i$ follows or replies to $j$ at least once; {repeated exchanges between the same pair are treated as a single tie.} We treat the network as an unweighted graph to focus on the structural boundaries of selective exposure (i.e., who connects to whom) rather than the frequency of interaction.
In the follower perspective, the neighbor set of $i$ is defined as $\mathcal{N}_i = \{j \mid (i, j) \in \mathcal{E}\}$, consisting of out-neighbors with whom $i$ interacts.
Conversely, in the leader perspective, the neighbor set is $\mathcal{N}_i = \{ j \mid (j, i) \in \mathcal{E}\}$, representing in-neighbors who interact with $i$.
For clarity, we exclude self-loops, meaning that an individual is never considered its own neighbor, i.e., $i \notin \mathcal{N}_i$ for all $i \in \mathcal{I}$.
Each individual $i \in \mathcal{I}$ holds an opinion $x_i \in [0,1]$ on a topic, where smaller values indicate opposition, larger values indicate support, and $x_i = 0.5$ reflects neutrality.

%In a BC-inspired model, 
{From a bounded confidence perspective,}
the opinion of the $i$-th individual (``Ego'') is compared with that of its neighbors in $ \mathcal{N}_i$. 
Denoting by $ \bar{x}_{\mathcal{N}_i} = \frac{1}{|\mathcal{N}_i|} \sum_{j \in \mathcal{N}_i } x_j $ the average opinion of Ego's neighbors, a useful quantity to measure is the opinion gap
\begin{equation}
g_i = x_i - \bar{x}_{\mathcal{N}_i}
\label{eq:opin-gap}
\end{equation}
which quantifies how far Ego's opinion is from the average opinion of its neighbors.

%To check whether the homophily in our data obeys distance-dependent rules, we conduct an ego-level monotonicity test relating opinion distance to local neighbor counts. For each Ego $i$ (with at least two neighbors), we compute the pairwise distances 
%\begin{equation}\label{equ:Distance}
%	\Delta_{ij} = |x_i - x_j|
%\end{equation}
%to its neighbors $j \in \mathcal{N}_i$, bin these distances, and plot the resulting neighbor counts as a function of distance. We then test whether the nonzero values in this histogram are strictly decreasing with distance; Egos satisfying this condition are labeled as distance-dependent (``DD'') while they are labeled ``NDD'' (non-distance-dependent) otherwise.
%A less stringent test consists of computing the Pearson correlation coefficient between opinion distance and neighbors counts. Negative correlations correspond to some level of distance-dependent decay; positive correlations do not.

Denote $ \alpha_i = \min_{j \in \mathcal{N}_i} x_j $ and $ \beta_i = \max_{j \in \mathcal{N}_{i}} x_j$. 
For each $i$, the interval $\left[\alpha_i,\beta_i\right]$ is termed the confidence interval of $i$, representing the observed span of opinions among Ego's neighbors, and its width
\begin{equation}\label{equ: Ci}
	c_i =\beta_i - \alpha_i 
\end{equation}
is referred to as the confidence range, quantifying the dispersion of opinions in Ego's neighborhood. Obviously, $ 0 \leqslant c_i \leqslant 1 $.
{For the analyses below, quantities involving $\alpha_i$, $\beta_i$, and $c_i$ are considered only for users with at least two neighbors, whereas the opinion gap and deviation measures can also be computed when a user has only one neighbor.}

A BC-inspired interpretation emphasizes two ingredients: a tendency to interact with similar others (homophily) and the existence of a finite interaction cutoff (boundedness). The confidence range alone is not sufficient to operationalize these features, so we introduce three additional indices quantifying the extent of opinion alignment within an individual's local neighborhood.

The first measure is the mean opinion deviation, defined for individual $i$ as
\begin{equation}\label{equ:MeanDeviation}
	\Delta_{\text{mean}}(i) 
	= \frac{1}{|\mathcal{N}_i|} \sum_{j \in \mathcal{N}_i} \left| x_i - x_j \right|
\end{equation}
%$\Delta_{\text{mean}}(i)$
%which captures the central tendency of Ego's neighborhood in opinion space. It quantifies the overall local alignment tendency, providing a more fine-grained measure of overall alignment than $ g_i$.
which quantifies {the average local alignment} of Ego's neighborhood in opinion space, providing a more fine-grained measure than $g_i$.

We further define the max opinion deviation of individual $i$ as
\begin{equation}\label{equ:MaxDeviation}
	\Delta_{\text{max}}(i) = \max_{j\in\mathcal{N}_i} \lvert x_i - x_j \rvert
\end{equation}
%This metric probes the boundedness assumption of the BC model.
which {quantifies the outer extent of Ego's neighborhood in opinion space.}
A smaller $\Delta_{\text{max}}(i)$ suggests that interactions are confined within a finite distance from Ego, consistent with a bounded interaction horizon.
% emphasized in BC models.

The third quantity we use is a binary index called the range inclusion condition, which states that individual $i$'s opinion lies within the minimum-to-maximum opinion confidence interval of its neighbors:
\begin{equation}\label{equ:RangeInclusion}
	\alpha_i \leq x_i \leq \beta_i.
\end{equation}
This condition checks for local consistency, verifying whether Ego 
%is within the ideological span of their neighbors, as expected by BC-inspired rules.
lies within {the observed opinion span of its neighbors.}

In addition, we need to characterize the potential asymmetry in the confidence interval with respect to the Ego's opinion. For that, denoting 
\begin{equation}
\label{eq:gamma-delta}
\gamma_i = x_i - \alpha_i \; \; \text{and } \;\; \delta_i = \beta_i - x_i,
\end{equation}
%the left and right confidence range, we can define the asymmetry index:
{we define the left and right offsets of the observed neighbor interval relative to Ego, which may be signed.
When $x_i\in[\alpha_i,\beta_i]$, these quantities coincide with nonnegative lower and upper confidence ranges; otherwise, one of them becomes negative, indicating that Ego lies outside the observed opinion span of its neighbors.
This motivates the asymmetry index}
\begin{equation}\label{equ:AsymmetryIndex} 
	s_i = (\delta_i - \gamma_i)/(\delta_i + \gamma_i)= (\delta_i - \gamma_i)/c_i. 
\end{equation} 
%A positive $s_i$ indicates greater tolerance for opinions higher (i.e., closer to 1) than Ego's own, while a negative $s_i$ suggests greater tolerance for opinions lower (i.e., closer to 0) than Ego's own.
A positive $s_i$ indicates {a larger extension of the neighbor span toward} higher opinions (i.e., closer to 1), while a negative $s_i$ indicates {a larger extension toward} lower opinions (i.e., closer to 0).

%We can aggregate the three individual-level BC conditions
We can aggregate the three individual-level quantities
in Eqs.~\eqref{equ:MeanDeviation}, \eqref{equ:MaxDeviation} and \eqref{equ:RangeInclusion} into three population-wide indicators.
The mean deviation rate, $R_1$, quantifies the fraction of individuals whose mean opinion deviation among empirical neighbors is smaller than that among randomized neighbors (see \nameref{subsec:DesignNull} for more details on the null model adopted):
\begin{equation}\label{equ:MeanDeviationRate}
	R_1 = \frac{1}{N} \sum_{i=1}^N \mathbf{1}\Bigl(\Delta_{\text{mean}}(i) < \Delta_{\text{mean}}^{\text{rand}}(i)\Bigr),
\end{equation}
where $\Delta_{\text{mean}}(i)$ and $\Delta_{\text{mean}}^{\text{rand}}(i)$ denote the mean opinion deviations for empirical and randomized neighbors, respectively, and $\mathbf{1}(\cdot)$ denotes the indicator function, returning 1 if the condition inside holds and 0 otherwise.
Thus, $R_1$ measures how often empirical neighborhoods are more centrally concentrated around Ego than expected from the null model.

The maximal deviation rate, $R_2$, is defined as
\begin{equation}\label{equ:MaxDeviationRate}
	R_2 = \frac{1}{N}\sum_{i=1}^N \mathbf{1}\Bigl(\Delta_{\text{max}}(i) < \Delta_{\text{max}}^{\text{rand}}(i)\Bigr),
\end{equation}
where $\Delta_{\text{max}}(i)$ and $\Delta_{\text{max}}^{\text{rand}}(i)$ are the maximal opinion deviations in the empirical and randomized cases, respectively.
It captures how often empirical neighborhoods exhibit a tighter outer boundary in opinion space than expected under randomization.

Finally, the range inclusion rate, $R_3$, quantifies instead the fraction of individuals whose opinions fall within the observed interval spanned by their neighbors:
\begin{equation}\label{equ:RangeInclusionRate}
	R_3 = \frac{1}{N} \sum_{i=1}^N \mathbf{1}\left( \alpha_i \leq x_i \leq  \beta_i \right).
\end{equation}
This provides a population-level summary of the local inclusion condition, quantifying how frequently users remain within the observed range of their peers.

\subsection*{Design of null models}
\label{subsec:DesignNull}

We construct two null models with increasing levels of constraint to benchmark the homophily content in our data. Both null models randomize interaction patterns while strictly preserving the relevant degree, i.e., neighbor count $\lvert \mathcal{N}_i\rvert$, of each individual, thereby controlling for the skewed activity distributions (as noted in \textit{SI} Fig.~S3).

In the first, denoted randomized null model, each individual $i$ is assigned $\lvert \mathcal{N}_i\rvert$ neighbors drawn randomly from the entire population.
This model destroys opinion-interaction correlations, serving as a baseline for unconstrained mixing. Comparisons against this model reveal the presence of general homophily induced solely by the global opinion distribution.

In the second, denoted range-based null model, neighbors are randomly selected from the population but constrained to fall within 
%a confidence range $c_{i}^\text{rand}$ matching the empirical width $c_i$.
{a candidate interval whose width is determined by} the empirical confidence range $c_i$.
This model is more stringent: it reshuffles identities while preserving both the neighbor count and 
%the localized interaction width 
{the scale of the localized interaction horizon} within the bounded opinion space $[0,1]$.
By thus accounting for geometric constraints and platform-specific structural constraints (e.g., ceiling effects \cite{sirbu2019algorithmic} or local confinement), it provides a conservative baseline for $R_1$, $R_2$, and $R_3$.
%Deviations beyond this baseline indicate residual non-random structure consistent with more concentrated interactions within the available range, rather than artifacts of boundary confinement alone.
Comparisons against this baseline help assess whether empirical neighborhoods remain more concentrated within the available range than would be expected from boundary confinement and other geometric constraints alone.
{A step-by-step pseudocode specification of the range-based null model is provided in \textit{SI} Section~S3.}

\subsection*{Social media datasets}
\label{subsec:Data}

We analyze three datasets from two influential online social platforms: Reddit and Twitter, each consisting of both opinion quantifications and directed interaction patterns.
Further details, data summaries, and a discussion on platform-specific interaction norms are provided in \textit{SI} Text and \textit{SI} Table~S1.

\noindent\textbf{\textit{Dataset \#1: Reddit-politics.}}
The Reddit dataset of \cite{pansanella2022change} spans 24 months of user interactions (May 2018--April 2020) in a politically focused subreddit, \href{https://www.reddit.com/r/politics/}{\texttt{r/politics}}, involving over $10^5$ users and covering the highly polarized mid-term period of the first Trump presidency.
User opinions are inferred at the post level via an LSTM classifier trained on partisan subreddits; monthly user scores are the average of their post scores, a step taken to reduce context-dependent noise and estimate the user's latent ideological stance in a more robust way, mapped onto a continuous scale ranging from 0 (anti-Trump) to 1 (pro-Trump), with 0.5 denoting neutrality.
Directed edges represent user replies within the same discussion thread.
Here we show our results for the first month (May 2018), while the results for some of the remaining months (June 2018--April 2020) are provided in \textit{SI} (except for Fig.~\ref{fig: case_2_Reddit_Population-level_satisfaction_rates_new}, where all 24 months are shown).

\noindent\textbf{\textit{Dataset \#2: Twitter-Covid.}}
The COVID-19 Twitter dataset captures users' opinion exchanges on pandemic-related topics through tweet replies.
Using the Twitter Academic Research API, we collected 337,258 Italian‐language tweets containing vaccination-related keywords or hashtags during the peak of the vaccination campaign debate in Italy (March 2021), a period marked by intense public division over health mandates.  
Tweets are labelled on a 5-point pro/anti-vaccine scale by a multilingual BERT model fine-tuned on 86,340 balanced training examples.
A user's overall opinion score is then computed as the average of the user's tweet-level sentiments, normalized to the interval $[0,1]$, where 0 means totally against vaccination and 1 means totally in favor.
Directed edges represent reply interactions between users.

\noindent\textbf{\textit{Dataset \#3: Twitter-contentious.}}
The second Twitter dataset focuses on ideological and political topics, covering six contentious issues (abortion, gun control, Obamacare, the US 2020 vice presidential debate, the US 2020 second presidential debate, and the US 2020 election day)
\cite{hohmann2023quantifying}.
User opinions are extracted from shared hyperlinks, quantified along a liberal-to-conservative spectrum using data from 
\href{https://www.mediabiasfactcheck.com}{mediabiasfactcheck.com}.
A user's overall opinion is computed as the average of the ideological scores of the domains they shared.
Data from \cite{hohmann2023quantifying} are linearly rescaled, mapping the opinion interval $[-1,1]$ of \cite{hohmann2023quantifying} into $[0,1]$, where 0 means the most liberal and 1 means the most conservative stances.
Directed edges are constructed from follower links. 
We showcase here the US 2020 second presidential debate dataset, while results for other topics are summarized in \textit{SI} (except for Fig.~\ref{fig: case_2_Reddit_Population-level_satisfaction_rates_new}, where all six topics are shown).

\subsection*{Statistical testing and effect sizes}\label{subsec:Stats}

We use the standard two‐sample $t$-test to determine whether the empirical distributions of confidence range and opinion deviation differ significantly from the randomized null models. 
Given the large sample sizes in our datasets (especially Reddit-politics and Twitter-Covid), even negligible differences could yield statistically significant $p$-values. 
Therefore, to assess the practical magnitude of these differences, we report Cohen's $d$ effect size, with $d \approx 0.2$ commonly interpreted as small, $d \approx 0.5$ as medium, and $d \approx 0.8$ as large.
This metric is important for distinguishing statistically significant but minor deviations (often dominated by the global opinion distribution) from substantive residual structure that is not explained by the null constraints and is consistent with selective neighborhood formation.

\section*{Acknowledgments}
This work was supported in part by grants from the Swedish Research Council (grants n. 2020-03701 and 2024-04772 to C.A.) and by the ELLIIT framework program at Link\"oping University.
The authors would like to thank M. Coscia and M. Hohmann for sharing the Twitter-contentious dataset and for discussions on the topic of the paper.

\section*{Data Availability}

The Reddit-politics dataset and the Twitter-contentious dataset used in this study were obtained from previously published work~\cite{pansanella2022change,hohmann2023quantifying}. 
The Twitter-Covid dataset constructed by the authors during this study will be made publicly available upon acceptance of the article; until that time, the dataset is available from the corresponding author upon reasonable request.  
Code supporting the findings of this study is also available from the corresponding author upon reasonable request.

\bibliographystyle{unsrt}
{\small
\bibliography{bibreply}
}

\end{document}

\maketitle

\section{The BC opinion dynamics model}
\label{sec:BC}

The Bounded Confidence (BC) model offers a foundational framework for modeling interactions constrained by opinion similarity, where individuals interact only with neighbors whose opinion differences fall within their confidence bounds.
In the literature, the BC principle is used mainly to represent opinion dynamics, as it provides an intuitive and plausible opinion update rule grounded in the homophily principle \cite{hegselmann2002opinion,bernardo2024bounded}. 
Formally, in the opinion dynamics model the opinion update rule is expressed as 
\begin{equation}\label{equ: BC_dynamics}
	x_{i}^{+} = x_i + \frac{1}{\lvert \mathcal{N}_{i}(\mathbf{x})\rvert} \sum_{j\in\mathcal{N}_i(\mathbf{x})} (x_j - x_i),\ \forall i\in\mathcal{I}
\end{equation}
where $x_{i}:=x_{i}(t) $ and $ x_{i}^{+}:=x_{i}(t+1)$ denote the opinions of individual $i$ at consecutive discrete time steps $t$ and $t+1$, respectively, and $ \mathbf{x}:=(x_1,x_2,\ldots,x_N)$ is the vector of all opinions.
The set $\mathcal{N}_{i}(\mathbf{x})$ contains neighbors whose opinions differ from $x_i$ by at most $\gamma_i$ below or $\delta_i$ above, i.e.,
\begin{equation*}
	\mathcal{N}_{i}(\mathbf{x}) := \{j \in \mathcal{I} \mid -\gamma_{i}\leq x_j - x_i \leq \delta_{i}\}.
\end{equation*}
Here, $\gamma_i \ge 0$ and $\delta_i \ge 0$ are individual-specific lower and upper confidence thresholds, respectively, capturing heterogeneity in tolerance. 
Under certain convergence criteria, BC dynamics evolves into consensus or clustered equilibria, the latter characterized by distinct opinion groups of like-minded individuals.
As mentioned in the main text, in this work we are not concerned with the temporal dynamics in Eq.~\eqref{equ: BC_dynamics}, but only with 
%the neighbor selection rule 
{the state-dependent neighborhood structure} implied by the BC principle, namely the interaction neighborhoods defined by $\mathcal{N}_{i}(\mathbf{x})$.
{In the empirical analyses of the main text, we reuse the notation $\gamma_i,\delta_i$ for the left and right offsets of the observed neighbor interval relative to Ego's opinion, which may thus be signed. These empirical quantities coincide with the usual nonnegative BC confidence thresholds when $x_i\in[\alpha_i,\beta_i]$.}

\section{Datasets details}
\label{sec:DataSummary}

In this section, we detail the data collection and preprocessing pipelines. For a discussion on how the distinct debate cultures and interaction norms of these platforms shape the resulting network topologies, please refer to Section~\ref{ssec:topologyAnalysis}.

\subsection{Dataset \#1: Reddit-politics}

This dataset is derived from the work of Pansanella et al.~\cite{pansanella2022change}, who analyzed user interactions in the politically focused subreddit of Reddit, \href{https://www.reddit.com/r/politics/}{\texttt{r/politics}}. The authors collected posts and comments spanning a 24-month window (May 2018 through April 2020) via the Pushshift API~\cite{baumgartner2020pushshift}.

The original Reddit data include posts and subsequent comments (``replies''), which form the directed interactions: when user $i$ replies to user $j$, there is a directed edge $(i,j)$.
In this work reply edges capture engagement/discussion and may include confrontational exchanges; accordingly, our reply-based results are interpreted as structural patterns of bounded engagement rather than direct evidence of interpersonal influence.
In total, more than 1 million unique users engaged in this subreddit, although typical monthly snapshots are on the order of 100,000--300,000 users, see Table~\ref{tab:DataSummary}.

Pansanella et al.\ treat the user-alignment task as a text classification problem. Each Reddit post is assigned a model ``leaning score'' in $[0,1]$, where 0 represents an anti-Trump (Democrat-leaning) position and 1 represents a pro-Trump (Republican-leaning) position,
using an LSTM neural network trained on data from known pro-Trump and anti-Trump subreddits~\cite{morini2021toward,morini2020capturing}. For each user in a given month, the authors average that user's post-level scores to produce a monthly user-level score.  
Combining these scores with the reply edges yields the final directed network of users with monthly ideology estimates.

\subsection{Dataset \#2: Twitter-Covid}

This dataset was assembled specifically for this study, hence we give more details on its compilation.
We collected posts from Twitter using the Academic Research API, which provided free access to public data from the platform's full archives (available from 2021 to 2023). To identify a relevant timeframe for data collection, we analyzed the volume of topic-specific tweets related to COVID-19 vaccination in the Italian Twitter-sphere throughout 2021. Based on this analysis, we selected March 2021 as the period of interest, during which discussions on COVID-19 vaccination were particularly prevalent.

To retrieve the relevant tweets, we employed Twarc2, a Python library that facilitates the collection of Twitter data in JSON format. We defined a query using the following set of topic-specific hashtags and keywords which were identified based on their prominence in vaccine-related discourse during 2021:

\begin{center}
	\begin{tabular}{c}
	     \#novaccino  \#iovaccino \#libertadiscelta  \#vaccinocovid19 \\ \#iononmivaccino  \#iomivaccino  \#novax  \#provax  \#iononsonounacavia  \#dittaturasanitaria \\
	     \#vienegiututto - \#pfizerdown - \#vaccino - \#vaccini - \#covid19 - \\
	     \#covid - \#coronavirus - \#vaccinazioni - \#vaccinocovid - vaccino covid-19 \\
	     vaccinazione covid-19 vaccino covid  vaccinazione vaccini \\
	\end{tabular}
\end{center}

When constructing the query, we restricted the language to Italian and excluded tweets from verified accounts, since their posting behavior can differ substantially from that of regular users. Additionally, we executed queries to collect replies, quotes, retweets, and likes associated with a subset of the dataset, which required separate treatment in the analysis.

The final dataset comprises tweets from 59,677 users, with a total of 337,258 tweets. We preprocessed the collected tweets by removing URLs, converting text to lowercase, and eliminating Twitter-specific special characters. However, we retained emojis, allowing the sentiment analysis tool we chose, BERT, to contextualize their usage within the dataset, as sarcasm and irony in social media discourse are often conveyed through non-standard emoji associations.

\paragraph{Sentiment analysis and model fine-tuning.}

In the sentiment analysis literature, several studies~\cite{munikar2019fine, d2019monitoring, ansari2021worldwide} have fine-tuned BERT models for fine-grained multi-class sentiment analysis using the SST-5 dataset. Previous efforts to establish a baseline dataset of Italian tweets include~\cite{polignano2019alberto}, where BERT was adapted for NLP tasks such as sentiment analysis in Italian. However, this model primarily supports binary sentiment classification.

To enable more nuanced sentiment classification, we adopted the approach proposed in~\cite{pota2021effective, alturayeif2021fine} by employing a version of BERT pretrained on plain text. This model is capable of capturing contextual meanings across various domains. We used a pretrained BERT-base-multilingual-uncased model, which consists of 12 layers with a hidden size of 768 neurons and was pre-trained on multilingual data. Unlike models specifically fine-tuned for social media text classification, we pre-labeled tweets into five sentiment classes. This model, trained on more than 500,000 product reviews in six languages, including Italian, predicts sentiment on a scale from 1 to 5 (where 1 = ``totally against vaccination'' and 5 = ``totally in favor''). Before fine-tuning, we balanced the dataset to ensure that each sentiment class contained an equal number of tweets. This step was essential to prevent class imbalance from skewing model performance. The final balanced dataset used for fine-tuning consisted of 86,340 tweets, with 17,268 tweets per class.

To validate the model's effectiveness in analyzing vaccine-related tweets, we first conducted a visual inspection of sample sentences. Then, we fine-tuned the BERT-base-multilingual-cased model, during which BERT adjusted all its parameters based on the labeled tweets, learning to classify new tweets accordingly. This methodology follows the approach outlined in \cite{pota2021effective}, which achieved state-of-the-art performance on the SENTIPOLC 2016 dataset \cite{sentipolc2016}, the most widely recognized dataset for sentiment analysis in Italian.

We performed fine-tuning k-fold validation for three epochs using a learning rate of $2 \cdot 10^{-5}$, a parameter found to be optimal for text classification on tweets, as suggested in \cite{sun2019fine}.

\paragraph{Building the dataset for the bounded confidence analysis.} 

As our sentiment analysis architecture has been trained on a fine-grained 5-classes dataset, its outcome is discrete. To associate to each of the tweets a real-valued sentiment in the interval $[0,1]$, we considered the scores attributed by the classifier before applying the last softmax layer. Namely, the continuous sentiment of each tweet has been constrained to belong to the interval defined by the class associated to the tweet by the classifier (for instance, $[0, 0.2]$ for tweets in class 1), while the specific value within this interval was computed based on the scores attributed by the classifier to the remainder of the classes.

The overall dataset we assembled consisted of 24,842 replies to 10,656 referenced tweets. 
From this dataset, we built the directed interaction graph $ \mathcal{G} = \{\mathcal{V}, \mathcal{E}\} $ according to the following procedure: the nodes in $\mathcal{V}$ are the union, without repetitions, of user IDs that replied to a tweet or that authored the referenced tweets; when node $i$ replied to node $j$, the edge $(i,j)$ belongs to the set $\mathcal{E}$. 
As in the Reddit reply network, reply edges here capture engagement/discussion (which may also be confrontational) rather than confirmed interpersonal influence; therefore, our reply-based results should be interpreted as structural patterns of bounded engagement.
To each node, aka user ID, we assigned a sentiment score through our BERT architecture, by averaging the sentiment scores of the tweets/replies produced by that user. 
Doing so we obtained a graph with 16,791 nodes (user IDs) and 23,053 edges (replies), as summarized in Table~\ref{tab:DataSummary}.

\subsection{Dataset \#3: Twitter-contentious}

This dataset follows the approach of Hohmann et al.~\cite{hohmann2023quantifying}, who examine six highly divisive political or ideological issues on Twitter: abortion, gun control, Obamacare, the US 2020 vice-presidential debate, the US 2020 second presidential debate, and the US 2020 election day.

The authors build a follower network by collecting each target user's follower relationships (up to a limit imposed by Twitter's API, yet retrieving all directed links without filtering for ideology or reciprocity). These connections are stored as directed edges $(i,j)$ if user $i$ follows user $j$,
and when restricting the graph to the topic-specific node sets used in our analysis, we observe zero reciprocity (no mutual follower pairs) across all six topics.
In each of the six topics, the authors retrieve the relevant tweet IDs from prior studies~\cite{cinelli2021echo,DVN/UCJUUZ_2021}, re-collect them, and filter users for a minimum tweeting threshold. This inclusion criterion ensures that the sampling relies strictly on activity participation (engagement with the topic) rather than specific ideological stances or connectivity patterns.

For each user, Hohmann et al.\ analyze URLs the user has shared. They map each news source or domain onto a continuous $[-1,1]$ interval for a ``liberal-to-conservative'' spectrum, using the third-party fact-checking site~\href{https://www.mediabiasfactcheck.com}{mediabiasfactcheck.com} to derive each source's ideological tilt. A user's opinion is computed as the average of the domain-level scores for all URLs that the user shares in tweets.
After building the follower graph, each user obtains an overall ideological score in $[-1,1]$, with $-1$ signifying the most liberal and $+1$ signifying the most conservative stances.
To uniform with the other datasets, in our work the $[-1, 1]$ interval is linearly mapped into $[0, 1]$.   
The final edge sets and node sets counts vary with the topic, see Table~\ref{tab:DataSummary}.

\subsection{Statistics of the datasets}

Table~\ref{tab:DataSummary} presents some basic features of the analyzed datasets across different time periods and discussion topics. 
It reports the total number of users engaged in interactions ($N$) and the total number of directed interactions ($\lvert\mathcal{E}\rvert$). 
Note that users with opinions but no observed interactions are excluded from $N$ in Table~\ref{tab:DataSummary}.
Additionally, it details user counts with at least one in-neighbor ($N_{\text{in},\geq 1}$) or out-neighbor ($N_{\text{out},\geq 1}$), as well as those with at least two in-neighbors ($N_{\text{in},\geq 2}$) or out-neighbors ($N_{\text{out},\geq 2}$).
In fact, while opinion-based analyses can be carried out for users with at least one neighbor, confidence range analyses require at least two neighbors.

\begin{table}[H]
	\centering
	\begin{threeparttable}
	\caption{\textbf{Summary statistics of the analyzed datasets.}}
	\label{tab:DataSummary}
	\small
	\begin{tabular}{l l r r r r r r r}
		\hline
		Media & Dataset & Time & \multicolumn{1}{c}{$N$} & \multicolumn{1}{c}{$\lvert\mathcal{E}\rvert$} & \multicolumn{1}{c}{$ N_{\text{in},\geq 1}$} & \multicolumn{1}{c}{$ N_{\text{in},\geq 2}$} & \multicolumn{1}{c}{$N_{\text{out},\geq 1}$} & \multicolumn{1}{c}{$N_{\text{out},\geq 2}$} \\
		\hline
		\rowcolor{RedditColor!20} Reddit & US politics & May 2018 & 129584 & 1229972 & 83545 & 57452 & 127392 & 76135 \\
		\rowcolor{RedditColor!20} &  & Jun 2018 & 150348 & 1391128 & 97355 & 66830 & 127392 & 88163 \\
		\rowcolor{RedditColor!20} &  & Jul 2018 & 150666 & 1394510 & 95996 & 65820 & 148023 & 87800 \\
		\rowcolor{RedditColor!20} &  & Aug 2018 & 151914 & 1361611 & 95219 & 64869 & 149351 & 87815 \\
		\rowcolor{RedditColor!20} &  & Sep 2018 & 162164 & 1512061 & 104297 & 72418 & 159580 & 96458 \\
		\rowcolor{RedditColor!20} &  & Oct 2018 & 169801 & 1432973 & 109224 & 74750 & 166842 & 98765 \\
		\rowcolor{RedditColor!20} &  & Nov 2018 & 193881 & 1543111 & 120038 & 81192 & 190338 & 109599 \\
		\rowcolor{RedditColor!20} &  & Dec 2018 & 167146 & 1257429 & 101117 & 67237 & 164415 & 92083 \\
		\rowcolor{RedditColor!20} &  & Jan 2019 & 229359 & 2002954 & 139770 & 95456 & 227108 & 131477 \\
		\rowcolor{RedditColor!20} &  & Feb 2019 & 212502 & 1660021 & 130183 & 89039 & 210482 & 120663 \\
		\rowcolor{RedditColor!20} &  & Mar 2019 & 204729 & 1582461 & 124360 & 84105 & 202554 & 115332 \\
		\rowcolor{RedditColor!20} &  & Apr 2019 & 166300 & 1279258 & 102106 & 68576 & 164526 & 93224 \\
		\rowcolor{RedditColor!20} &  & May 2019 & 164198 & 1259475 & 100439 & 67294 & 162305 & 91411 \\
		\rowcolor{RedditColor!20} &  & Jun 2019 & 173254 & 1271842 & 104985 & 70577 & 171576 & 96306 \\
		\rowcolor{RedditColor!20} &  & Jul 2019 & 201006 & 1542275 & 121360 & 81907 & 198939 & 112662 \\
		\rowcolor{RedditColor!20} &  & Aug 2019 & 196398 & 1451403 & 115763 & 77067 & 194388 & 108561 \\
		\rowcolor{RedditColor!20} &  & Sep 2019 & 188025 & 1414031 & 111048 & 74435 & 186288 & 104116 \\
		\rowcolor{RedditColor!20} &  & Oct 2019 & 205876 & 1704680 & 122350 & 82774 & 203873 & 114961 \\
		\rowcolor{RedditColor!20} &  & Nov 2019 & 198058 & 1565306 & 117344 & 79333 & 196237 & 110717 \\
		\rowcolor{RedditColor!20} &  & Dec 2019 & 211110 & 1509982 & 123986 & 83229 & 209724 & 115654 \\
		\rowcolor{RedditColor!20} &  & Jan 2020 & 213088 & 1642999 & 127430 & 85049 & 211154 & 118011 \\
		\rowcolor{RedditColor!20} &  & Feb 2020 & 247054 & 2109739 & 150448 & 102704 & 244748 & 141690 \\
		\rowcolor{RedditColor!20} &  & Mar 2020 & 306901 & 2496664 & 180233 & 121227 & 304607 & 172627 \\
		\rowcolor{RedditColor!20} &  & Apr 2020 & 254941 & 1536783 & 141815 & 90983 & 252794 & 134319 \\
		\hline
		\rowcolor{TwitterCovid!60} Twitter & COVID-19 & 2021 & 16791 & 23053 & 4160 & 1920 & 13658 & 4009 \\
		\rowcolor{TwitterOther!40} & Abortion & 2015--2016 & 2211 & 55328 & 1823 & 1654 & 2202 & 2105 \\
		\rowcolor{TwitterOther!40} & Gun control & 2015--2016 & 1092 & 22471 & 873 & 763 & 1081 & 989 \\
		\rowcolor{TwitterOther!40} & Obamacare & 2015--2016 & 204 & 1377 & 153 & 126 & 193 & 162 \\
		\rowcolor{TwitterOther!40} & US VP debate & Oct 2020 & 5407 & 116249 & 3416 & 2808 & 5379 & 4784 \\
		\rowcolor{TwitterOther!40} & US SP debate & Oct 2020 & 4697 & 94497 & 2888 & 2324 & 4671 & 4067 \\
		\rowcolor{TwitterOther!40} & US election day & Nov 2020 & 3965 & 40443 & 2546 & 2004 & 3900 & 3228 \\
		\hline
	\end{tabular}
	\begin{tablenotes}
		\item $N$ denotes the total number of users, 
		$\lvert\mathcal{E}\rvert$ denotes the number of directed interactions,
		$N_{\text{in},\geq 1}$ and $N_{\text{in},\geq 2}$ denote the number of users with at least one, resp. two, in-neighbors, and
		$N_{\text{out},\geq 1}$, $N_{\text{out},\geq 2}$ the number of users with at least one, resp. two, out-neighbors.
	\end{tablenotes}
	\end{threeparttable}
\end{table}

\subsection{{Confidence range summaries}}
\label{ssec:ConfidenceRangeSummary}

{
Since the confidence range plays a central role in the bounded confidence interpretation of the empirical interaction neighborhoods, we report here a compact summary of its distribution across datasets and perspectives in Table~\ref{tab:ConfidenceRangeSummary}.
For each dataset and perspective, we consider users with at least two neighbors and summarize the empirical confidence range $c_i$ through its median and interquartile range (IQR).}

\begin{table}[H]
	\centering
	\begin{threeparttable}
		\caption{\textbf{{Empirical confidence range summaries across datasets and perspectives.}}}
		\label{tab:ConfidenceRangeSummary}
		\small
		\begin{tabular}{l l l r r r r r r}
			\hline
			& & & \multicolumn{3}{c}{Leader perspective} & \multicolumn{3}{c}{Follower perspective} \\
			\cline{4-9}
			Media & Dataset & \multicolumn{1}{r}{Month/topic} & \multicolumn{1}{c}{$N_{\text{in},\geq 2}$} & \multicolumn{1}{c}{$\mathrm{med}(c_i)$} & \multicolumn{1}{c}{$\mathrm{IQR}(c_i)$} & \multicolumn{1}{c}{$N_{\text{out},\geq 2}$} & \multicolumn{1}{c}{$\mathrm{med}(c_i)$} & \multicolumn{1}{c}{$\mathrm{IQR}(c_i)$} \\
			\hline
			
			\rowcolor{RedditColor!20} Reddit & US politics & \multicolumn{1}{r}{May 2018} & 57452 & 0.27 & 0.31 & 76135 & 0.33 & 0.36 \\
			\rowcolor{RedditColor!20} &  & \multicolumn{1}{r}{Jun 2018} & 66830 & 0.26 & 0.30 & 88163 & 0.32 & 0.35 \\
			\rowcolor{RedditColor!20} &  & \multicolumn{1}{r}{Jul 2018} & 65820 & 0.26 & 0.31 & 87800 & 0.34 & 0.36 \\
			\rowcolor{RedditColor!20} &  & \multicolumn{1}{r}{Aug 2018} & 64869 & 0.26 & 0.31 & 87815 & 0.32 & 0.34 \\
			\rowcolor{RedditColor!20} &  & \multicolumn{1}{r}{Sep 2018} & 72418 & 0.26 & 0.30 & 96458 & 0.34 & 0.36 \\
			\rowcolor{RedditColor!20} &  & \multicolumn{1}{r}{Oct 2018} & 74750 & 0.26 & 0.31 & 98765 & 0.34 & 0.36 \\
			\rowcolor{RedditColor!20} &  & \multicolumn{1}{r}{Nov 2018} & 81192 & 0.27 & 0.31 & 109599 & 0.31 & 0.35 \\
			\rowcolor{RedditColor!20} &  & \multicolumn{1}{r}{Dec 2018} & 67237 & 0.27 & 0.31 & 92083 & 0.31 & 0.34 \\
			\rowcolor{RedditColor!20} &  & \multicolumn{1}{r}{Jan 2019} & 95456 & 0.26 & 0.31 & 131477 & 0.31 & 0.34 \\
			\rowcolor{RedditColor!20} &  & \multicolumn{1}{r}{Feb 2019} & 89039 & 0.26 & 0.30 & 120663 & 0.34 & 0.36 \\
			\rowcolor{RedditColor!20} &  & \multicolumn{1}{r}{Mar 2019} & 84105 & 0.26 & 0.30 & 115332 & 0.33 & 0.35 \\
			\rowcolor{RedditColor!20} &  & \multicolumn{1}{r}{Apr 2019} & 68576 & 0.26 & 0.30 & 93224 & 0.33 & 0.34 \\
			\rowcolor{RedditColor!20} &  & \multicolumn{1}{r}{May 2019} & 67294 & 0.27 & 0.31 & 91411 & 0.35 & 0.35 \\
			\rowcolor{RedditColor!20} &  & \multicolumn{1}{r}{Jun 2019} & 70577 & 0.26 & 0.30 & 96306 & 0.34 & 0.34 \\
			\rowcolor{RedditColor!20} &  & \multicolumn{1}{r}{Jul 2019} & 81907 & 0.26 & 0.30 & 112662 & 0.33 & 0.34 \\
			\rowcolor{RedditColor!20} &  & \multicolumn{1}{r}{Aug 2019} & 77067 & 0.26 & 0.30 & 108561 & 0.32 & 0.33 \\
			\rowcolor{RedditColor!20} &  & \multicolumn{1}{r}{Sep 2019} & 74435 & 0.26 & 0.31 & 104116 & 0.33 & 0.35 \\
			\rowcolor{RedditColor!20} &  & \multicolumn{1}{r}{Oct 2019} & 82774 & 0.26 & 0.30 & 114961 & 0.33 & 0.35 \\
			\rowcolor{RedditColor!20} &  & \multicolumn{1}{r}{Nov 2019} & 79333 & 0.26 & 0.30 & 110717 & 0.35 & 0.36 \\
			\rowcolor{RedditColor!20} &  & \multicolumn{1}{r}{Dec 2019} & 83229 & 0.26 & 0.30 & 115654 & 0.36 & 0.35 \\
			\rowcolor{RedditColor!20} &  & \multicolumn{1}{r}{Jan 2020} & 85049 & 0.26 & 0.29 & 118011 & 0.33 & 0.34 \\
			\rowcolor{RedditColor!20} &  & \multicolumn{1}{r}{Feb 2020} & 102704 & 0.25 & 0.29 & 141690 & 0.30 & 0.32 \\
			\rowcolor{RedditColor!20} &  & \multicolumn{1}{r}{Mar 2020} & 121227 & 0.24 & 0.27 & 172627 & 0.29 & 0.31 \\
			\rowcolor{RedditColor!20} &  & \multicolumn{1}{r}{Apr 2020} & 90983 & 0.26 & 0.29 & 134319 & 0.33 & 0.33 \\
			
			\hline
			
			\rowcolor{TwitterCovid!60} Twitter & COVID-19 & Vaccine & 1920 & 0.37 & 0.47 & 4009 & 0.19 & 0.22 \\
			\rowcolor{TwitterOther!40} & Contentious & Abortion & 1654 & 0.54 & 0.42 & 2105 & 0.56 & 0.36 \\
			\rowcolor{TwitterOther!40} &  & Gun control & 763 & 0.45 & 0.39 & 989 & 0.43 & 0.36 \\
			\rowcolor{TwitterOther!40} &  & Obamacare & 126 & 0.52 & 0.32 & 162 & 0.52 & 0.29 \\
			\rowcolor{TwitterOther!40} &  & US VP debate & 2808 & 0.34 & 0.33 & 4784 & 0.35 & 0.20 \\
			\rowcolor{TwitterOther!40} &  & US SP debate & 2324 & 0.30 & 0.33 & 4067 & 0.29 & 0.26 \\
			\rowcolor{TwitterOther!40} &  & US election day & 2004 & 0.36 & 0.37 & 3228 & 0.36 & 0.39 \\

			\hline
		\end{tabular}
		\begin{tablenotes}
			\item {$N_{\text{in},\geq 2}$ and $N_{\text{out},\geq 2}$ denote, respectively, the numbers of users with at least two in-neighbors (leader perspective) and at least two out-neighbors (follower perspective). The empirical confidence range is defined as $c_i=\beta_i-\alpha_i$, where $\alpha_i=\min_{j\in\mathcal{N}_i}x_j$ and $\beta_i=\max_{j\in\mathcal{N}_i}x_j$ are the minimum and maximum opinions observed in Ego's neighborhood.}
		\end{tablenotes}
	\end{threeparttable}
\end{table}

\section{{Range-based null model}}
\label{sec:RangeBasedNullModel}

{For the interval-based analyses in the main text, we adopt a range-based null model that preserves the empirical neighborhood size and employs the observed confidence range to define a feasible randomized interaction horizon. Algorithm~\ref{alg:RangeBasedNull} gives the corresponding procedure for a generic Ego $i$.
Consider an Ego $i$ with opinion $x_i$ and empirical neighbor-opinion multiset $\mathcal{O}_i=\{x_j\mid j\in\mathcal{N}_i\}$, with size $n_i=|\mathcal{O}_i|$. Let $c_i=\max(\mathcal{O}_i)-\min(\mathcal{O}_i)$ be the empirical confidence range of Ego $i$.}

\begin{algorithm}[H]
	\caption{{\textbf{Range-based null model for Ego $\bm i$}}}
	\label{alg:RangeBasedNull}
	\small
	\begin{algorithmic}[1]
		\Require empirical opinion list $\{x_k\}_{k=1}^N\subset[0,1]$, Ego opinion $x_i$, empirical neighborhood $\mathcal{O}_i$, neighborhood size $n_i$, confidence range $c_i$
		\Ensure randomized neighborhood $\widetilde{\mathcal{O}}_i$
		
		\If{$n_i=1$}
		\State Draw one opinion uniformly at random from the empirical opinion list (excluding $x_i$)
		\State Set $\widetilde{\mathcal{O}}_i$ equal to the resulting singleton
		\Else
		\State $\mathrm{flag}\gets 0$
		\While{$\mathrm{flag}=0$}
		\State Draw a center opinion $x_{\mathrm c}$ uniformly at random from the empirical opinion list
		\State Set
		\[
		a \gets x_{\mathrm c}-\frac{c_i}{2}, \qquad
		b \gets x_{\mathrm c}+\frac{c_i}{2}
		\]
		\If{$a<0$}
		\State $b \gets b-a$, \quad $a \gets 0$
		\ElsIf{$b>1$}
		\State $a \gets a-(b-1)$, \quad $b \gets 1$
		\EndIf
		\State Construct the candidate multiset
		\[
		\mathcal{C}_i=\{x_k \mid a\le x_k\le b,\ k\neq i\}
		\]
		\If{$|\mathcal{C}_i|\ge n_i$}
		\State Sort $\mathcal{C}_i$ in increasing order
		\State Include its minimum and maximum opinions
		\State Sample the remaining $n_i-2$ opinions uniformly without replacement from the interior of the sorted multiset
		\State Set the resulting multiset as $\widetilde{\mathcal{O}}_i$
		\State $\mathrm{flag}\gets 1$
		\EndIf
		\EndWhile
		\EndIf
	\end{algorithmic}
\end{algorithm}

{The construction preserves the empirical neighborhood size $n_i$ and the total span $c_i$ (but not the empirical left/right asymmetry). In the current implementation, $R_3$ is evaluated only for users with at least two neighbors, while for $R_1$ and $R_2$ the case $n_i=1$ is handled by the one-point randomization in Algorithm~\ref{alg:RangeBasedNull}.}

%\section{Correlation between opinion distance and interaction frequency}
\section{{Diagnostics of distance-dependent decay}}
\label{sec:DiagnosticsDD}

%Recall from the main paper that distance-dependent opinion interaction models typically require the frequency of user-user interactions to decrease with the opinion distance.
{To complement the interval-based analyses in the main text, we report here two simple diagnostics of how neighbor counts vary with opinion distance.
}

%For each Ego with at least two neighbors, we subdivide the opinion space into bins, count the neighbors of each user in each bin, and calculate the Pearson correlation coefficient between neighbor count and (binned) opinion distance. Fig.~\ref{fig:case_2_fig_correlation} shows that these coefficients are negative in around 80\% of the cases. This suggests that a general homophilic tendency exists for the majority of users: they are less likely to interact with partners as opinion distance increases.
For each Ego with at least two neighbors, we subdivide the opinion space into bins, count the neighbors falling in each bin, and calculate two quantities: (i) the Pearson correlation coefficient between neighbor count and binned opinion distance, and {(ii) a stricter monotonicity criterion, under which an Ego is labeled distance-dependent (``DD'') if the nonzero binned neighbor counts strictly decrease with distance, and non-distance-dependent (``NDD'') otherwise.
}

\begin{figure}[htb!]
	\centering
	\includegraphics[width=0.95\linewidth]{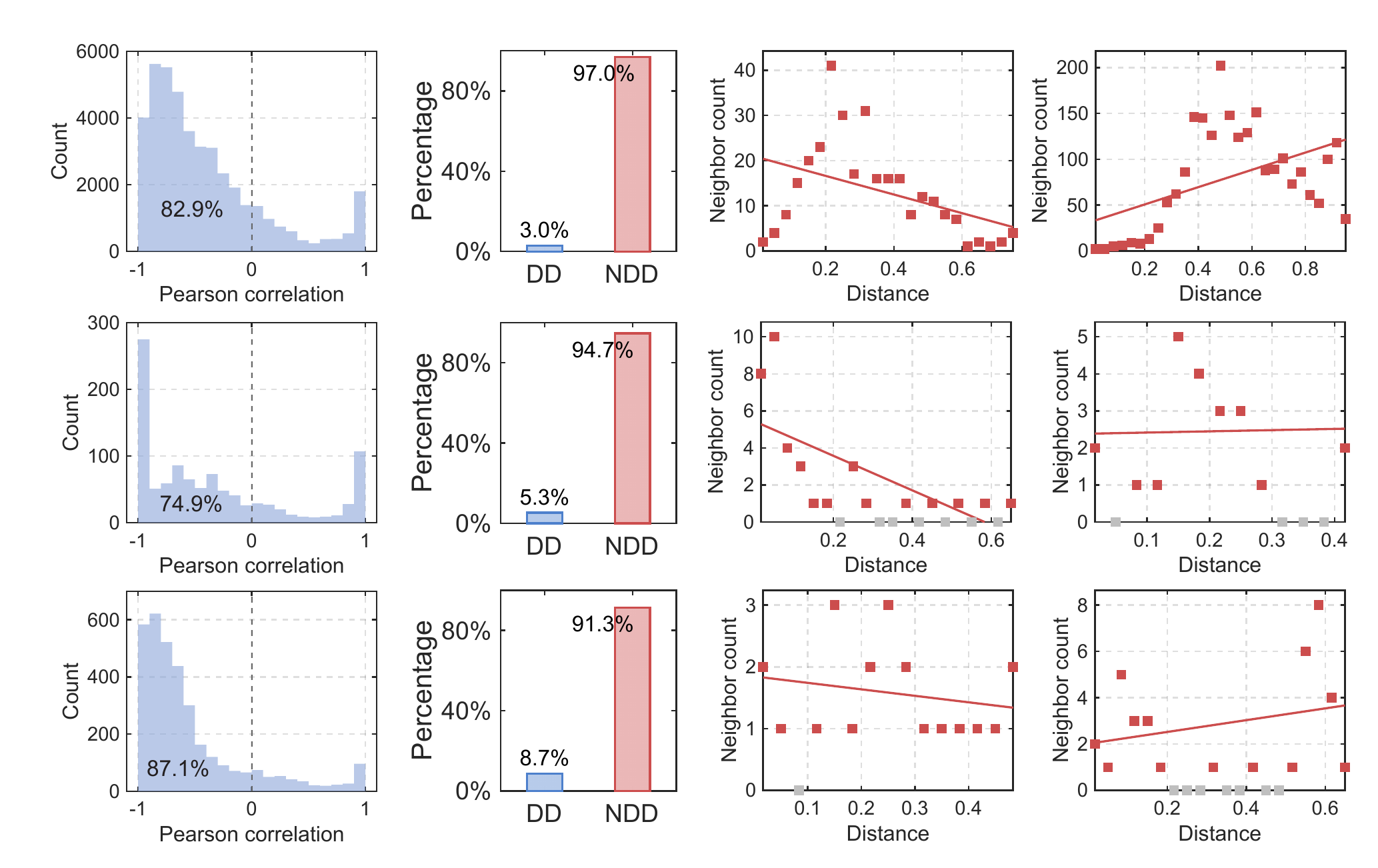}
	\caption{\textbf{{Distance-dependent decay diagnostics} across the three datasets.} Rows 1$\div$3: Reddit-politics, Twitter-Covid, Twitter-contentious. Column 1: Distributions of Pearson correlation coefficients. Column 2: Fractions of distance-dependent (``DD'') and non-distance-dependent (``NDD'') Egos according to the strict monotonicity criterion. Columns 3 and 4: Examples of users with negative and positive correlations, respectively. Gray markers indicate bins with zero neighbor counts; red lines show least-squares fits to nonzero bins.}
	\label{fig:case_2_fig_correlation}
\end{figure}

{Fig.~\ref{fig:case_2_fig_correlation} shows that Pearson correlations are negative in around 75--87\% of the cases. This suggests a broad homophilic tendency across all three datasets: for the majority of users, neighbor counts tend to decrease as opinion distance increases. By contrast, strict monotonicity is observed only in a small fraction of Egos (3.0\% for Reddit-politics, 5.3\% for Twitter-Covid, and 8.7\% for Twitter-contentious), indicating that these interaction profiles are rarely well described by a simple monotone decay.
}
{Pearson correlation here provides only a coarse summary of the overall trend, whereas the DD/NDD classification is sensitive to the binning of opinion distance and is interpreted as a bin-dependent diagnostic rather than a precise model test.}

However, relying solely on linear correlation or slope estimation comes with limitations due to the high variance in interaction patterns.
While a negative correlation is compatible with a distance-decay tendency, it does not guarantee robust estimation of a slope or boundary, especially when interaction profiles exhibit irregularities such as intermediate peaks (confrontational ties) or sparsity.
As the two right columns of panels in Fig.~\ref{fig:case_2_fig_correlation} show, in some cases, a least-squares fit could be used to provide a reasonable estimate of the slope and of the interaction boundary, but in others (even with negative correlation) the estimate becomes unreliable due to the non-smooth nature of the empirical data.
The examples shown here are selected to illustrate this heterogeneity. For each dataset we manually select two Egos among higher-degree users (to avoid overly sparse bins and improve visual interpretability): one with a negative correlation coefficient and one with a positive correlation coefficient.
For positive correlation, a distance-decay interpretation (and any boundary inferred from it) is not supported by the least-squares trend.

\section{Interaction preference test}

\begin{figure}[htb!]
  \centering
  \includegraphics[width=0.8\textwidth]{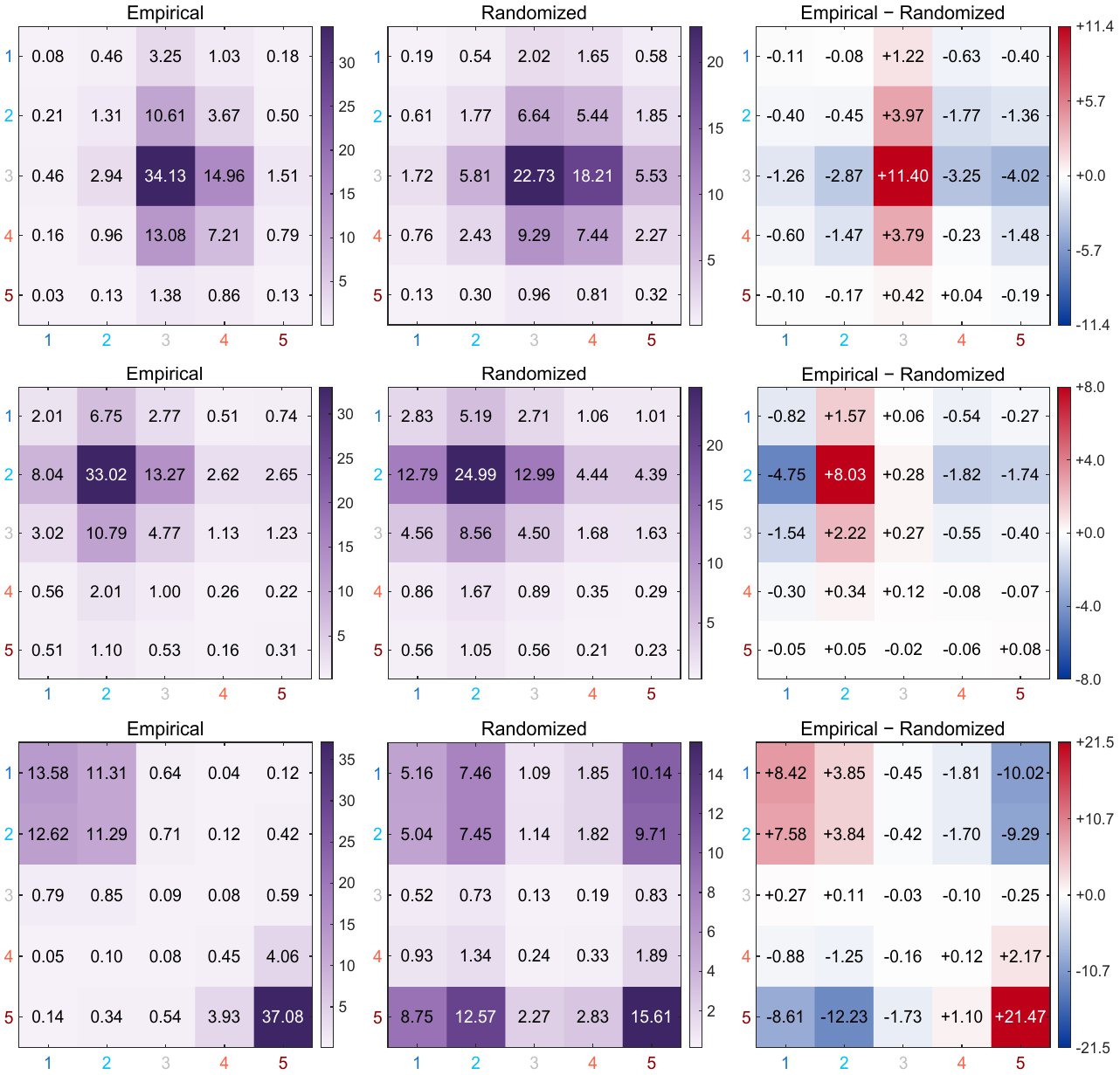}
  \caption{\textbf{Interaction patterns across opinion bins in three datasets.} 
  	Rows 1$\div$3: Reddit-politics, Twitter-Covid, Twitter-contentious. 
  	Columns 1$\div$3: empirical network, range-based randomized baseline (averaged over 20 trials), and difference (Empirical$-$Randomized), where red indicates excess over the null and blue indicates deficit. In Columns 1 and 2, each $(i,j)$ cell gives the percentage of interactions from opinion bin $i$ to bin $j$, normalized by the total interactions. {Column 3 reports the corresponding percentage-point differences.}}
  \label{fig:heatmap_all}
\end{figure}

As noted in the main text, all three datasets exhibit interaction patterns that deviate systematically from the randomized baseline (i.e., the range-based null model), which controls for the marginal opinion distributions.
In Fig.~\ref{fig:heatmap_all}, the left column reproduces the empirical heat maps of Fig.~5 in the main paper, the middle column shows the corresponding randomized baseline, and the right column shows the residual (Empirical$-$Randomized). 
In the reply-based networks, the central bins 
%attract more interactions than expected, indicating center-seeking behavior.
receive more interactions than expected under the randomized baseline. 
In the follow-based network, %extremes reinforce within-camp ties while suppressing cross-camp links.
{interactions are more concentrated within extreme camps and less frequent across camps} than expected under the randomized baseline.

\section{Analysis of the topology of the interaction graphs}
\label{ssec:topologyAnalysis}

Basic statistical observations indicate that the user-user interaction networks analyzed in the three datasets are generally sparse and weakly connected. In the reply-based networks (datasets \#1 and \#2), multiple weakly connected components exist, whereas the follow-based network (dataset \#3) forms a single weakly connected component. 
Moreover, in the topic-specific follow graphs of dataset \#3 we observe zero reciprocity, i.e., no users that follow each other, whereas the reply-based graphs (datasets \#1 and \#2) do contain bidirectional reply edges.
Note that this is a property of the induced subgraphs restricted to topic-participating users and the corresponding observed follower edge lists, as defined in \cite{hohmann2023quantifying}, and should not be interpreted as a general property of follower networks on Twitter.

We measured five node‐level indicators of centrality in our directed interaction networks: in‐degree, out‐degree~\cite{bonacich1972factoring}, local clustering coefficient~\cite{newman2006modularity}, Katz centrality~\cite{katz1953new}, and eigenvector centrality~\cite{bonacich1972factoring,bonacich1987power}.

In‐degree and out‐degree count respectively how many users comment on (or follow) a given node, and how many users the node comments on (or follows). Both metrics apply to directed networks and capture local ``popularity'' versus ``activity.'' 
We also compute a local clustering coefficient using a standard directed formulation, which measures how often a node's neighbors themselves interact.
For directed graphs, this local perspective can highlight small tightly knit subgroups (including potential echo chambers) and local bridging patterns.

Katz centrality and eigenvector centrality both capture global structural features of the networks: the Katz index leverages a damping rule to account for longer paths, while the eigenvector centrality highlights connections to other high‐scoring nodes.
Since our primary goal is to assess each user's overall prominence rather than local directionality, we symmetrize the directed network into an undirected form and then examine how each user's total degree correlates with these global centralities.

\begin{figure}[htb!]
	\centering
	\includegraphics[width=\linewidth]{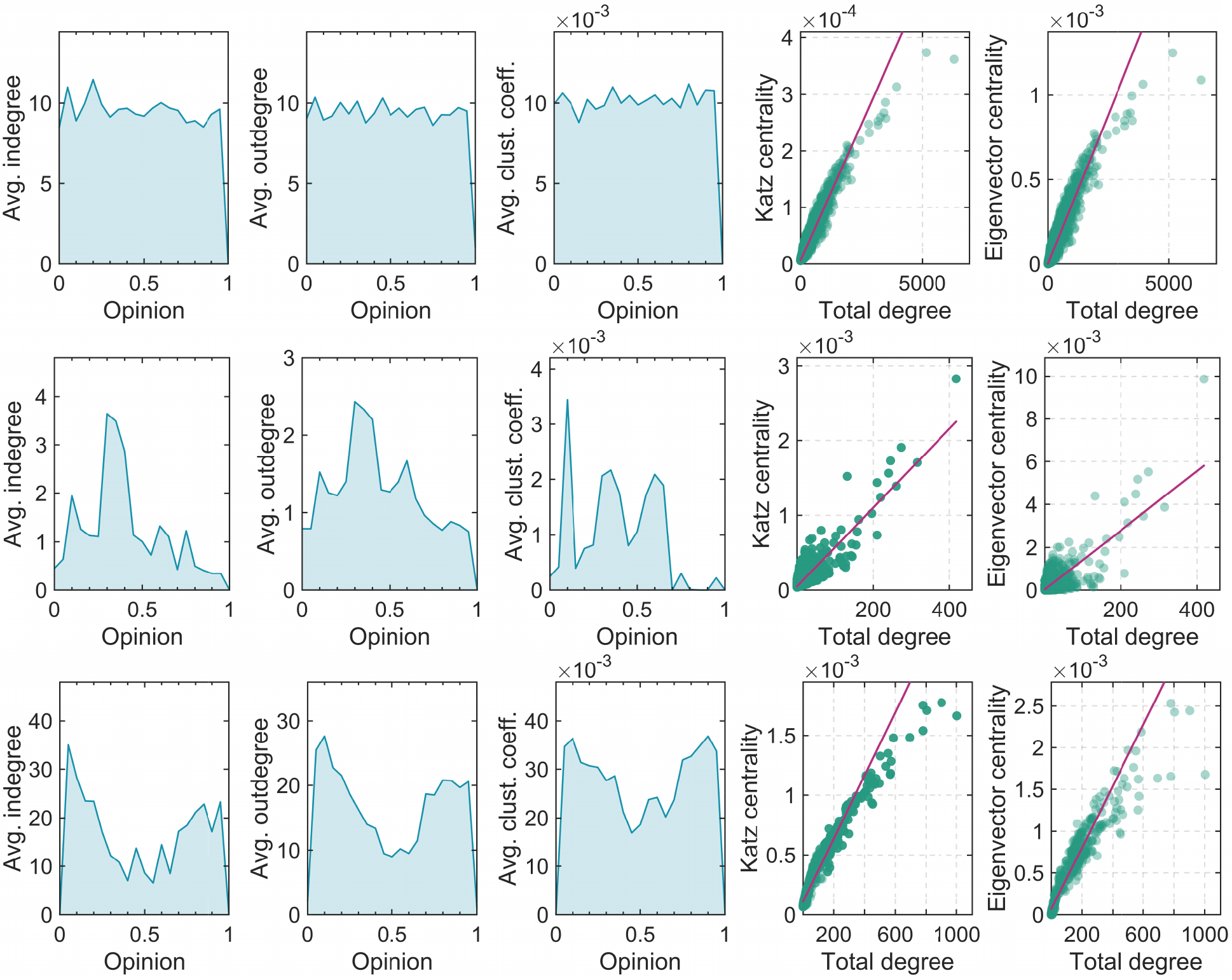}
	\caption{\textbf{Topological analysis of the interaction networks.} Top row: May~2018 of Reddit-politics.  Middle row: Twitter-Covid. Bottom row: Twitter-contentious US 2020 second presidential debate. Columns 1$\div$3: Distributions of the average in-degree, out-degree, and directed local clustering coefficient across different opinion intervals. Columns 4 and 5: Scatter plots between the total degree of users versus their Katz centrality and eigenvector centrality, respectively, with a purple linear fit highlighting the correlation.}
	\label{fig: centrality}
\end{figure}

As shown in Fig.~\ref{fig: centrality}, Reddit-politics shows a relatively ``flat'' pattern of average in/out‐degree and clustering coefficient across the opinion spectrum, whereas the two Twitter datasets exhibit a shape that somewhat resembles the opinion distribution.
In particular, it is clear from the correlation plots in the two rightmost columns of Fig.~\ref{fig: centrality} that individuals with high interaction frequency tend to be more important in the network.

There are several factors that can provide a potential explanation for the topological properties we see in Fig.~\ref{fig: centrality}.

\begin{description}
    \item[Platform culture and interaction norms:] 
    Reddit-politics is essentially one large forum in which users interact under one common thread, \texttt{r/politics}. The moderation style and wide, relatively centralized audience can yield more uniform participation patterns: Even if a user's stance is slightly left or right, they may still reply or be replied to at about the same frequency.  
	Twitter is organized around ``mentions'' and ``follows'' across a myriad of personal accounts. By design, users on Twitter-Covid or Twitter-contentious topics often self‐select into narrower, more homophilic sub‐groups. 
	Strongly opinionated accounts (whether pro or anti) tend to attract more followers or replies from like‐minded individuals, amplifying in‐degree/out-degree in the extreme opinion segments.

	\item[Topic specifics:] 
	The Reddit-politics dataset covers an entire subreddit about US politics: everyone's politics are channeled into the same ``room,'' so to speak. 
	While polarized discussions do occur, the single‐subforum structure can mix moderate voices and ensure at least some cross‐group exposure.  
	The Twitter-Covid and Twitter-contentious sets are narrower fora, focusing on particularly contentious issues, e.g., strong pro/anti positions on vaccination, abortion, gun control, etc. This can heighten the effect of ``echo chambers,'' so that extremely opinionated users do, in fact, get substantially higher degrees.
	
	\item[User distribution vs.\ interaction:]
	On Reddit, even if a user is extremely left or right, it may be exposed to and engage with replies from a broad base of participants. Meanwhile, many moderately positioned users can appear in each monthly snapshot. The net effect is a more uniform (even ``flatter'') average in/out‐degree curve across the opinion spectrum.  
	On Twitter, a strongly opinionated account may more effectively cultivate an ideological following, leading to large in‐degree in the extremes. Similarly, it may actively reply or retweet only within that ideological bubble, inflating out‐degree in those same opinion segments.
\end{description}

In short, the difference arises from a combination of platform design, topic selection, and user behavior. Reddit-politics has a single shared forum that somewhat ``flattens'' degree patterns by mixing participants. The Twitter datasets tend to concentrate on contentious issues, which fosters highly polarized or multipolar usage patterns and thus bimodal or multi-peaked in/out‐degree curves that mirror the underlying opinion extremes.

\section{Follower-perspective results on remaining datasets}
\label{sec:Follower-perspective results}

\begin{figure}[H]
	\centering
	\includegraphics[width=\linewidth]{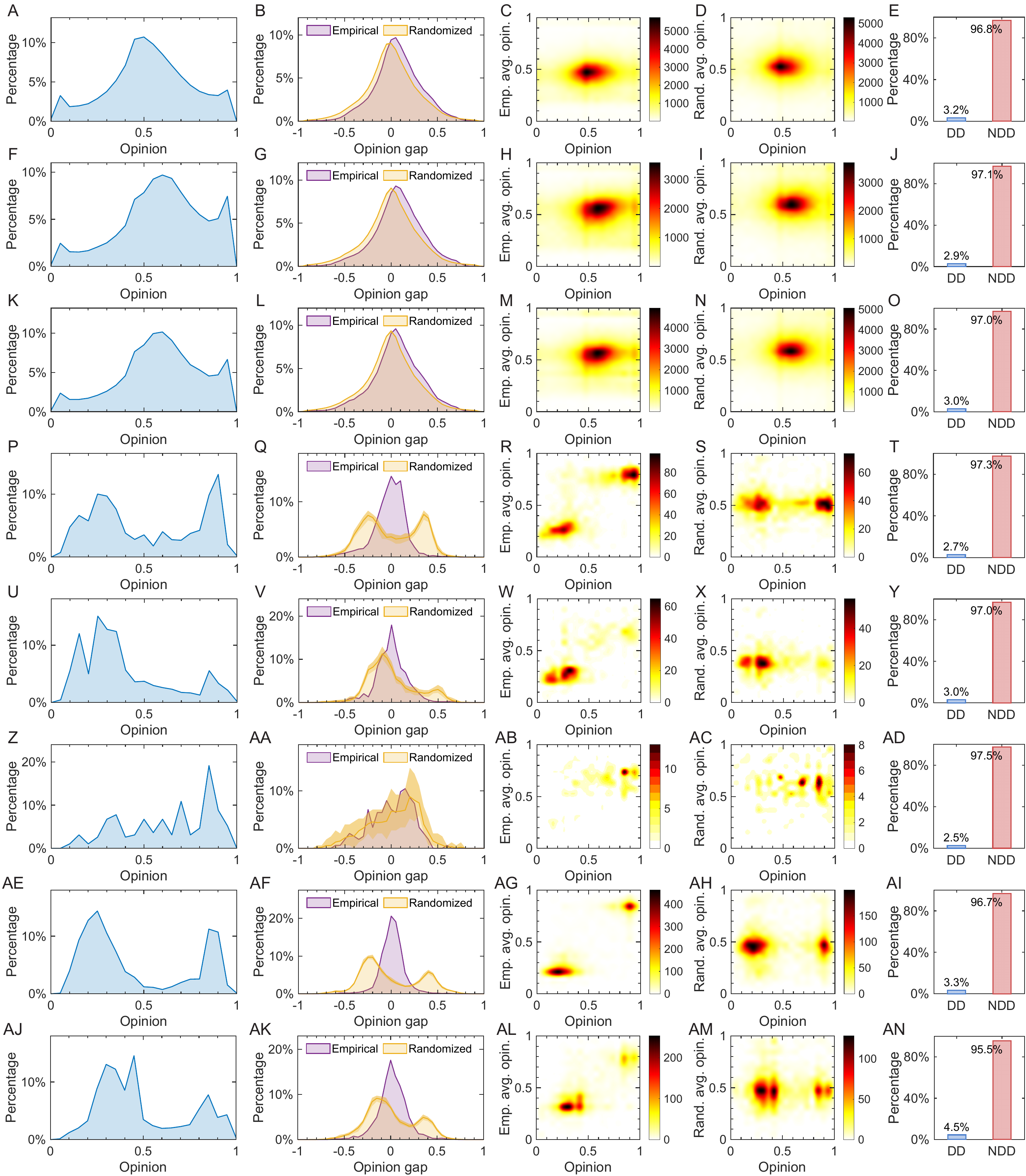}
	\caption{\textbf{Opinion distributions and {average-neighbor opinion comparison} across three extra months of the Reddit-politics dataset and the remaining five Twitter-contentious datasets.} Rows 1$\div$3: Reddit-politics dataset, Nov.~2018, May~2019, Nov.~2019. Rows 4$\div$8: Twitter-contentious Abortion, Gun control, Obamacare, US 2020 vice presidential debate, US 2020 election day. Column 1: Individual opinion distributions. Column 2: Opinion gap distributions. Column 3: Joint distributions of individual opinions and the average opinions of their empirical neighbors. Column 4: Corresponding joint distributions computed with randomized neighbor connections. {For reference, Column 5 reports the DD/NDD diagnostic based on the strict monotonicity criterion given in Section~\ref{sec:DiagnosticsDD}.}}
	\label{fig: case_2_opin_distribution_remaining}
\end{figure}

\begin{figure}[H]
	\centering
	\includegraphics[width=\linewidth]{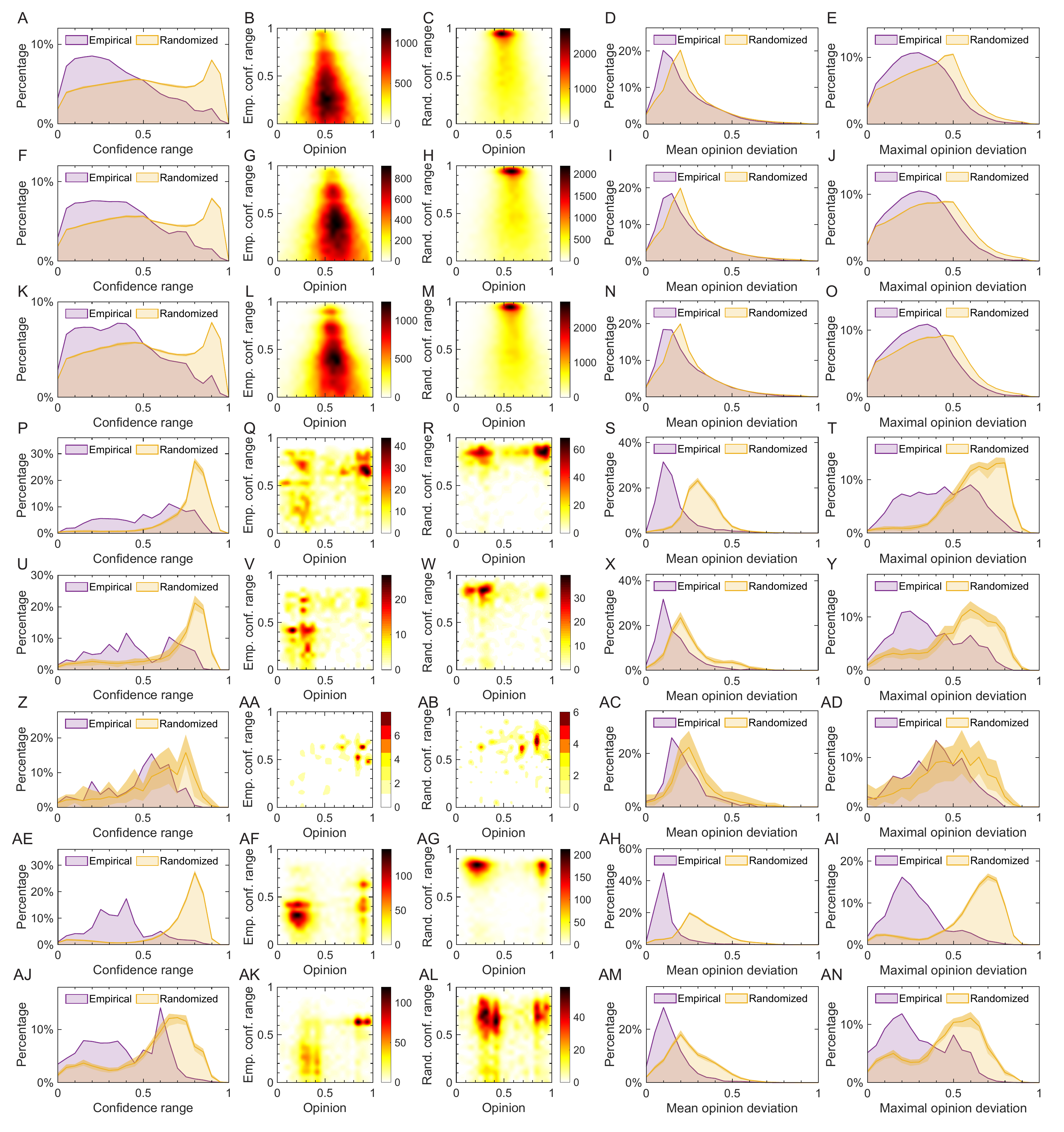}
	\caption{\textbf{Distributions of confidence range, mean deviation, and maximal deviation across three extra months of the Reddit-politics dataset and the remaining five Twitter-contentious datasets.} Rows 1$\div$3: Reddit-politics dataset, Nov. 2018, May 2019, Nov. 2019. Rows 4$\div$8: Twitter-contentious Abortion, Gun control, Obamacare, US 2020 vice presidential debate, US 2020 election day. Column 1: Individual confidence range distributions. Column 2: Joint distributions of individual opinions and their empirical confidence ranges. Column 3: Corresponding joint distributions using randomized confidence ranges. Column 4: Mean deviation distributions. Column 5: Maximal deviation distributions.}
	\label{fig: case_2_conf_range_distribution_remaining}
\end{figure}

\begin{figure}[H]
	\centering
	\includegraphics[width=0.7\linewidth]{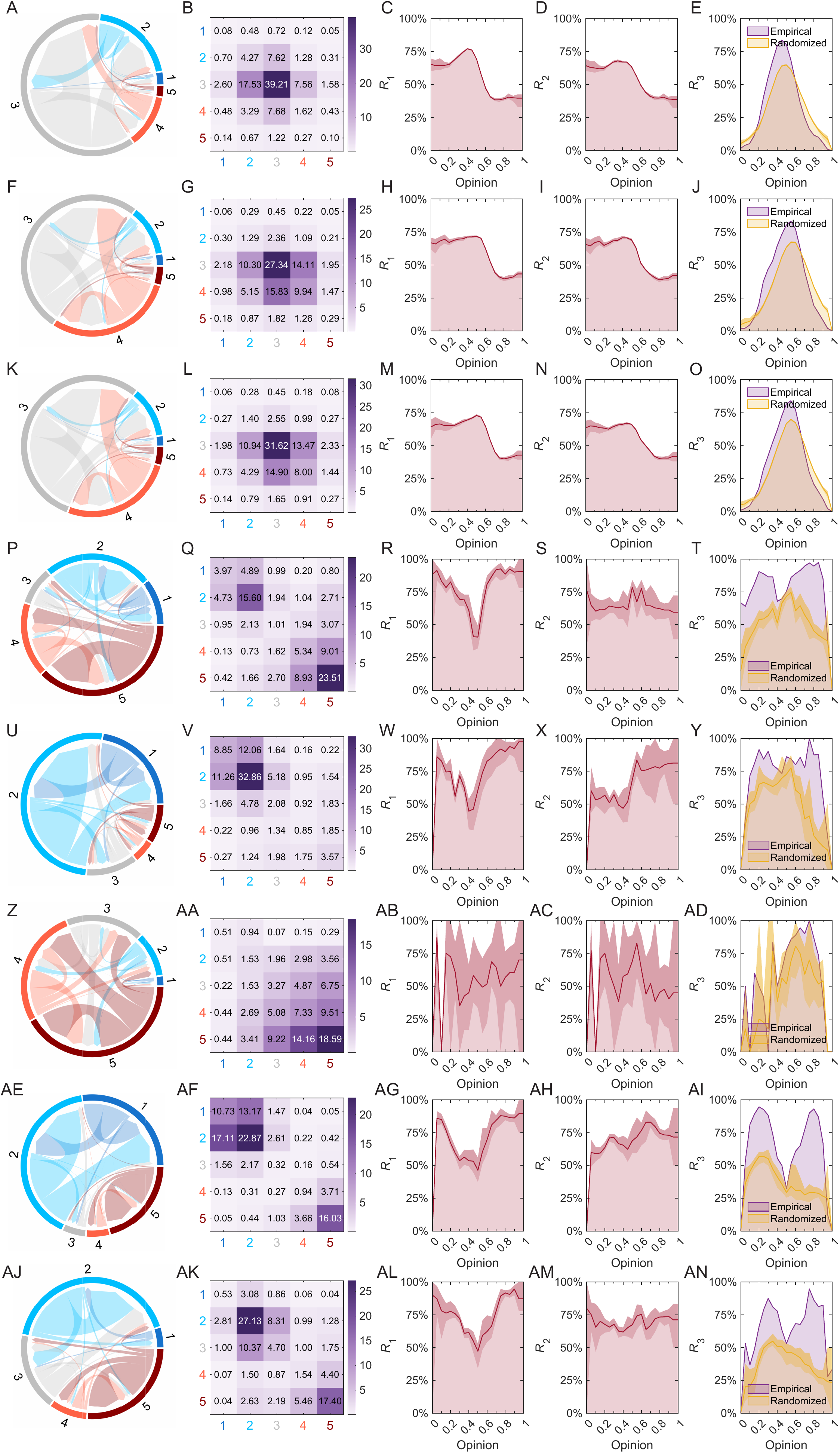}
	\caption{\textbf{Interaction patterns and {rates of interval-based conditions} across partition-based opinion intervals in three extra months of the Reddit-politics dataset and the remaining five Twitter-contentious datasets.} Rows 1$\div$3: Reddit-politics dataset, Nov. 2018, May 2019, Nov. 2019. Rows 4$\div$8: Twitter-contentious Abortion, Gun control, Obamacare, US 2020 vice presidential debate, US 2020 election day. Column 1: Chord diagrams illustrate the connectivity both within and between the groups. Column 2: Heatmaps represent the interaction percentages between opinion intervals. Columns 3$\div$5: {Rates for} mean deviation $R_1$, maximal deviation $R_2$, and range inclusion $R_3$ over 20 range-based randomized trials.}
	\label{fig: case_2_Individual-level_satisfaction_rates_new_remaining}
\end{figure}

\begin{figure}[H]
	\centering
	\includegraphics[width=\linewidth]{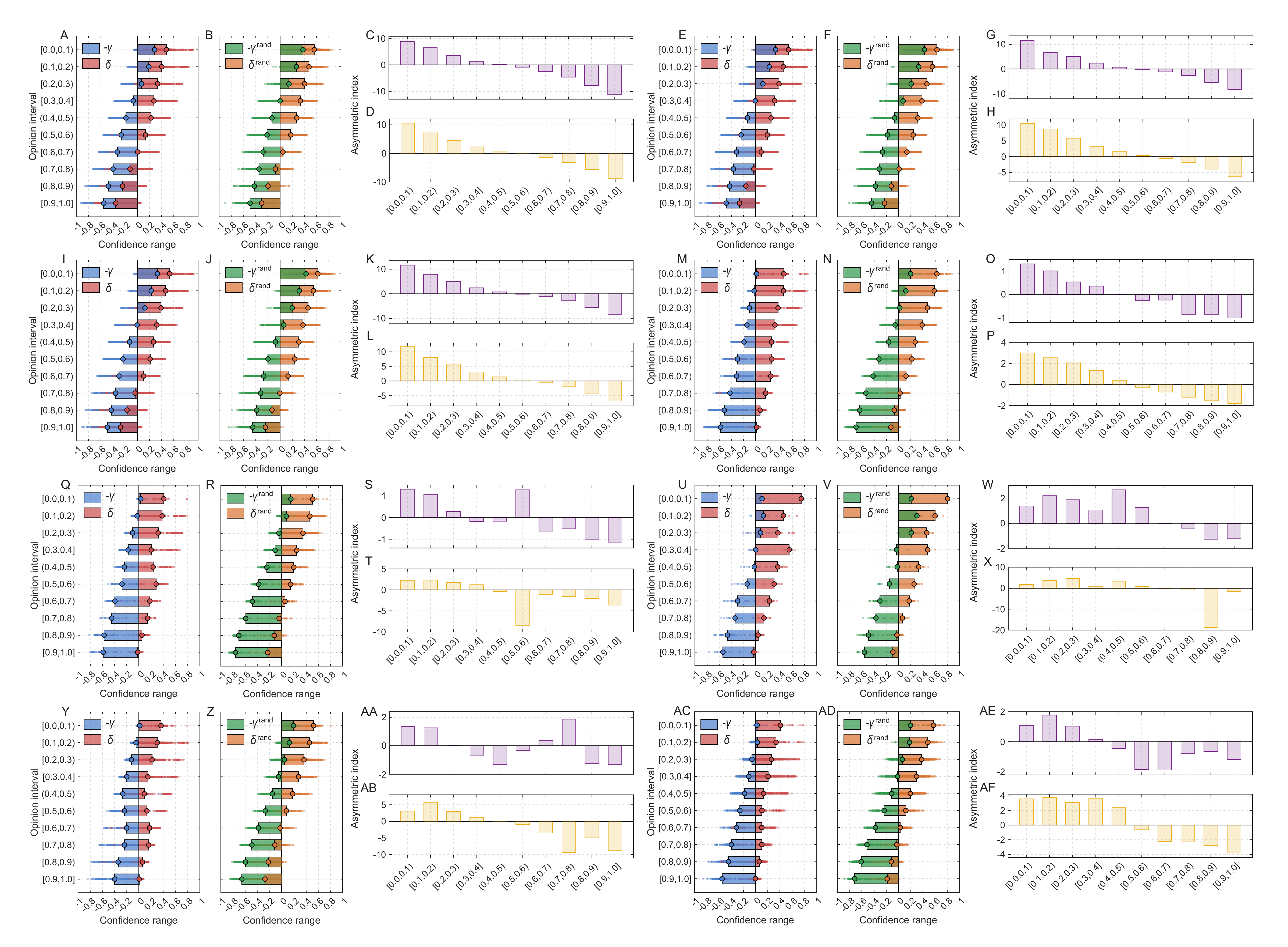}
	\caption{\textbf{Asymmetric {neighborhood spans} and opinion-dependent tolerance patterns across three extra months of the Reddit-politics dataset and the remaining five Twitter-contentious datasets.} Columns 1, 2, 4, 5: Distributions of negated left {offsets} ($-\gamma_i$) and right {offsets} ($\delta_i$) across different opinion intervals for Reddit-politics dataset, Nov. 2018 (A, B), May 2019 (E, F), Nov. 2019 (I, J), and Twitter-contentious Abortion (M, N), Gun control (Q, R), Obamacare (U, V), US 2020 vice presidential debate (Y, Z), and US 2020 election day (AC, AD). Columns 1 and 4 present empirical data, while Columns 2 and 5 show their corresponding range-based null models. Columns 3 and 6: Asymmetry index $s_i$ across different opinion intervals in empirical data (C, G, K, O, S, W, AA, AE), and the corresponding asymmetry index $s_i$ derived from range-based null models (D, H, L, P, T, X, AB, AF).}
	\label{fig: case_2_fig_add_LeftRightC_new_remaining}
\end{figure}

As mentioned in the main text, the Reddit-politics dataset spans 24 months and the Twitter-contentious six different topics. 
In this section we repeat our analysis for (some of) the months/topics not reported in the main manuscript.
In particular, in Figs.~\ref{fig: case_2_opin_distribution_remaining}--\ref{fig: case_2_fig_add_LeftRightC_new_remaining} we consider three additional months of the Reddit-politics data and the remaining five topics of the Twitter-contentious data, reporting the same analysis as in Figs.~2, 3, 5, and 6 of the main paper, still  under the follower scenario.

\section{Comparative results under the leader perspective}
\label{sec:Leader-perspective results}

\begin{figure}[H]
	\centering
	\includegraphics[width=\linewidth]{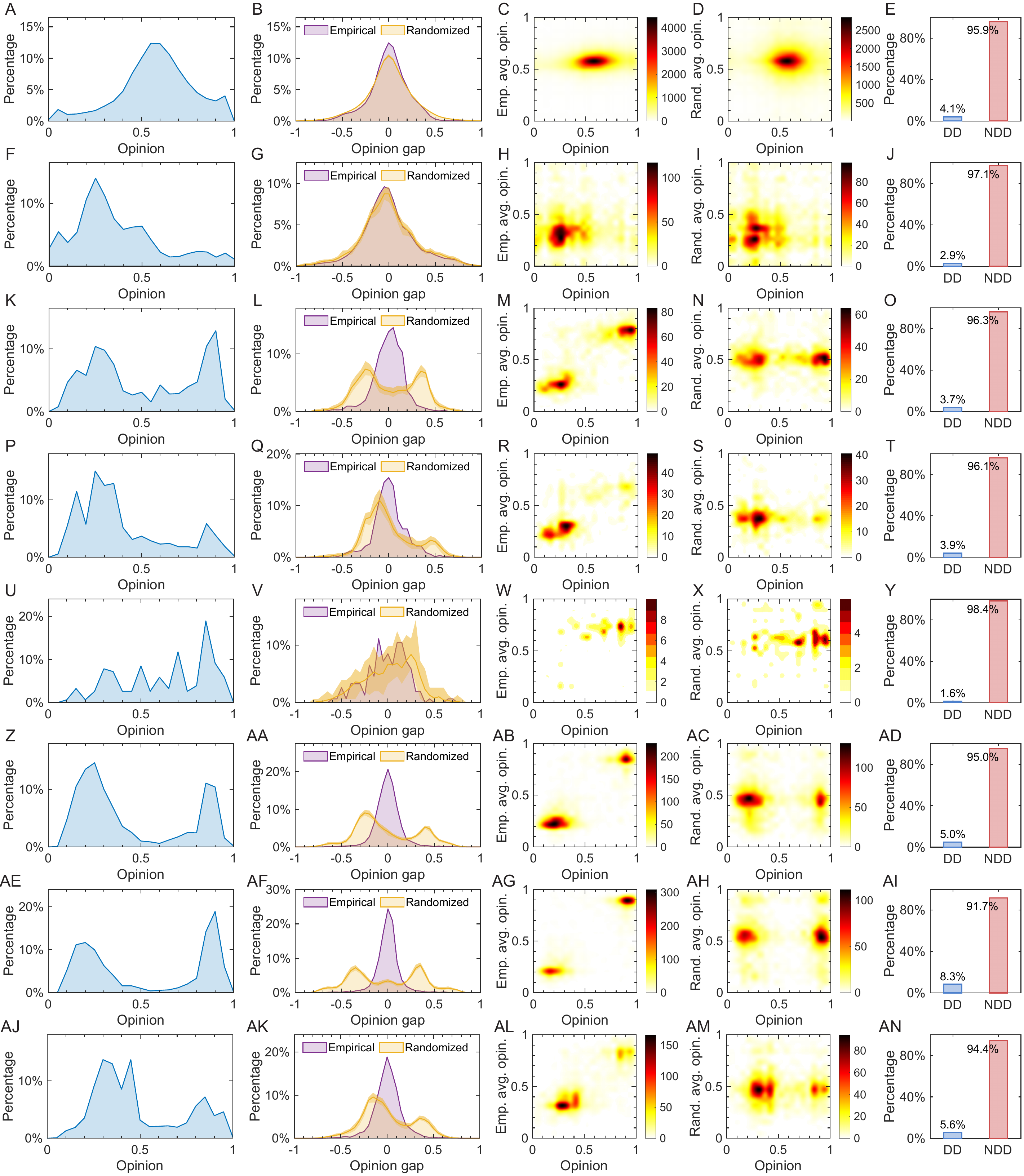}
	\caption{\textbf{Opinion distributions and {average-neighbor opinion comparison} across the first month of the Reddit-politics dataset and all Twitter datasets in the leader perspective.} Row 1: May~2018 of Reddit-politics. Row 2: Twitter-Covid. Rows 3$\div$8: Twitter-contentious Abortion, Gun control, Obamacare, US 2020 vice presidential debate, US 2020 second presidential debate, US 2020 election day. Column 1: Individual opinion distributions. Column 2: Opinion gap distributions. Column 3: Joint distributions of individual opinions and the average opinions of their empirical neighbors. Column 4: Corresponding joint distributions computed with randomized neighbor connections. {For reference, Column 5 reports the DD/NDD diagnostic based on the strict monotonicity criterion given in Section~\ref{sec:DiagnosticsDD}.}}
	\label{fig: case_1_fig_one_all}
\end{figure}

\begin{figure}[H]
	\centering
	\includegraphics[width=\linewidth]{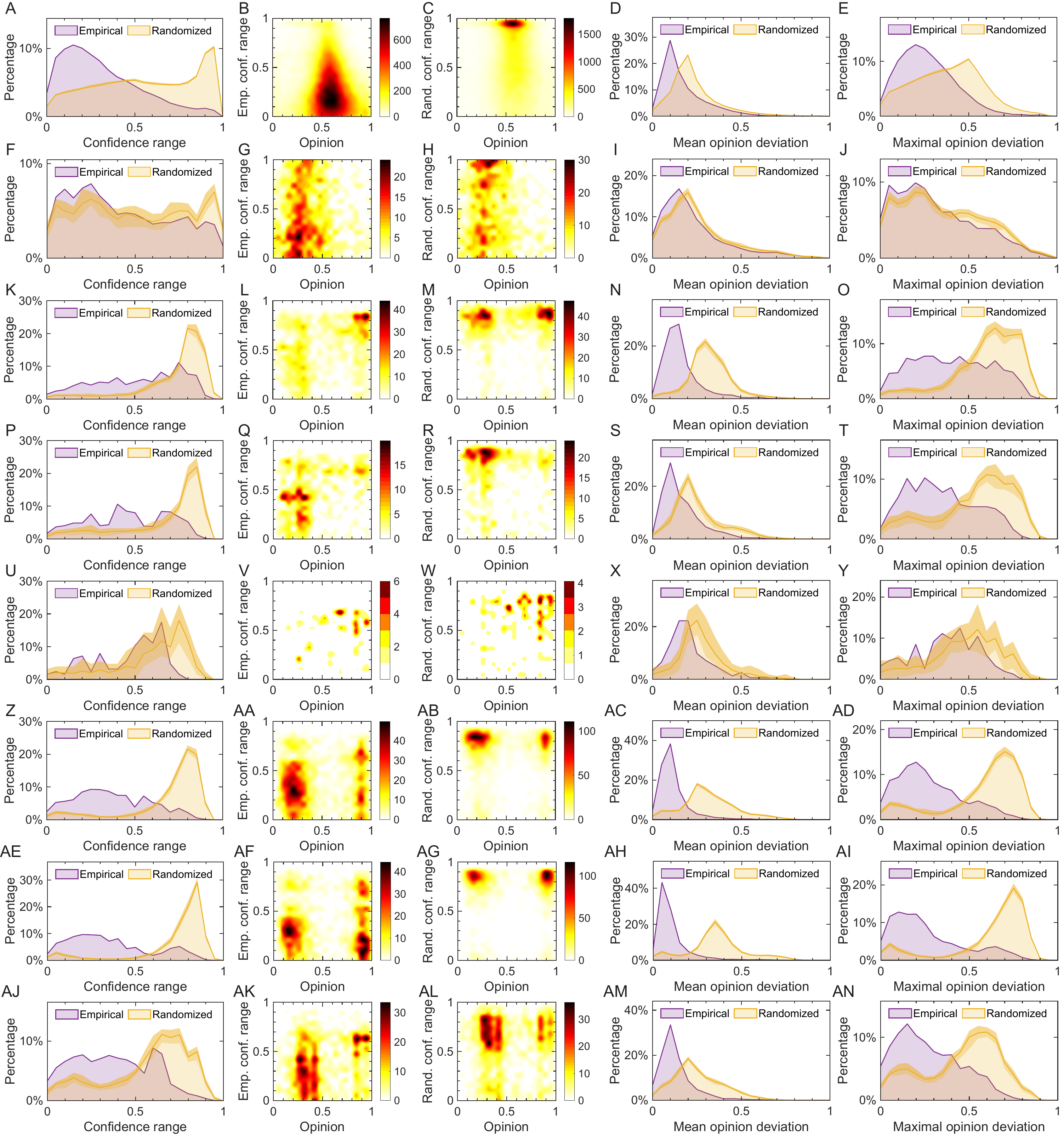}
	\caption{\textbf{Distributions of confidence range, mean deviation, and maximal deviation across the first month of the Reddit-politics dataset and all Twitter datasets in the leader perspective.} Row 1: May 2018 of Reddit--politics. Row 2: Twitter-Covid. Rows 3$\div$8: Twitter-contentious Abortion, Gun control, Obamacare, US 2020 vice presidential debate, US 2020 second presidential debate, US 2020 election day. Column 1: Individual confidence range distributions. Column 2: Joint distributions of individual opinions and their empirical confidence ranges. Column 3: Corresponding joint distributions using randomized confidence ranges. Column 4: Mean deviation distributions. Column 5: Maximal deviation distributions.}
	\label{fig: case_1_fig_two_all}
\end{figure}

\begin{figure}[H]
	\centering
	\includegraphics[width=0.9\linewidth]{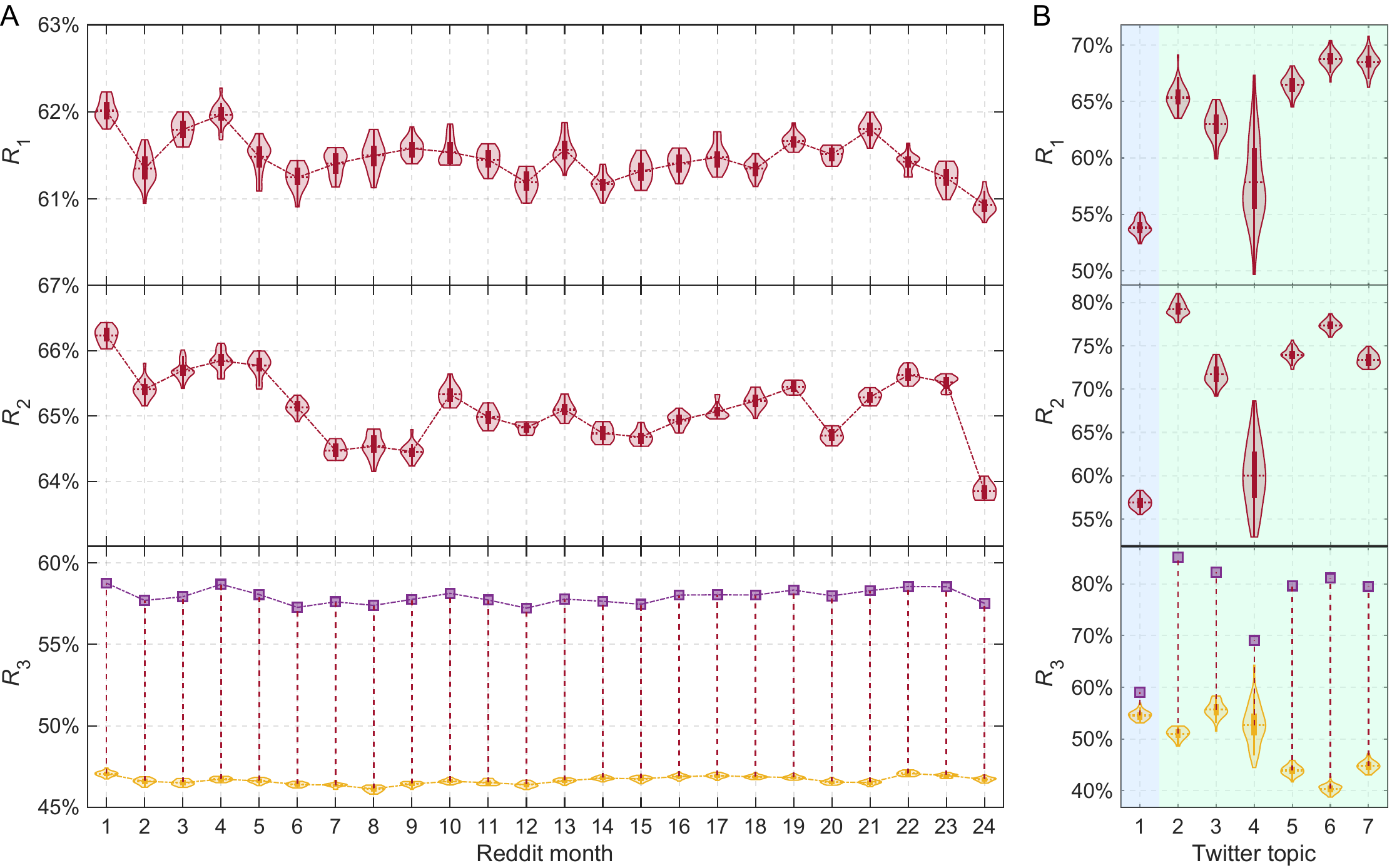}
	\caption{\textbf{{Population-level rates of interval-based conditions} in Reddit and Twitter datasets in the leader perspective.}
		(A)~Longitudinal trends for $R_1$, $R_2$, and $R_3$ over 24 months in the Reddit-politics data.
		(B)~Comparisons of these rates across multiple Twitter topics.  
		The first dataset (labeled ``1'', Twitter-Covid) is reply-based, while the remaining six datasets (labeled ``2''--``7'', Twitter-contentious topics) are follow-based. The violin plots in the first two rows show how $ R_1 $ and $ R_2 $ evolve over 20 random instances of the null model, while the last row compares the empirical and randomized $ R_3 $ (the latter, in yellow, over repeated instances represented as violin plots).}
	\label{fig: case_1_fig_five_Reddit_Twitter}
\end{figure}

\begin{figure}[H]
	\centering
	\includegraphics[width=\linewidth]{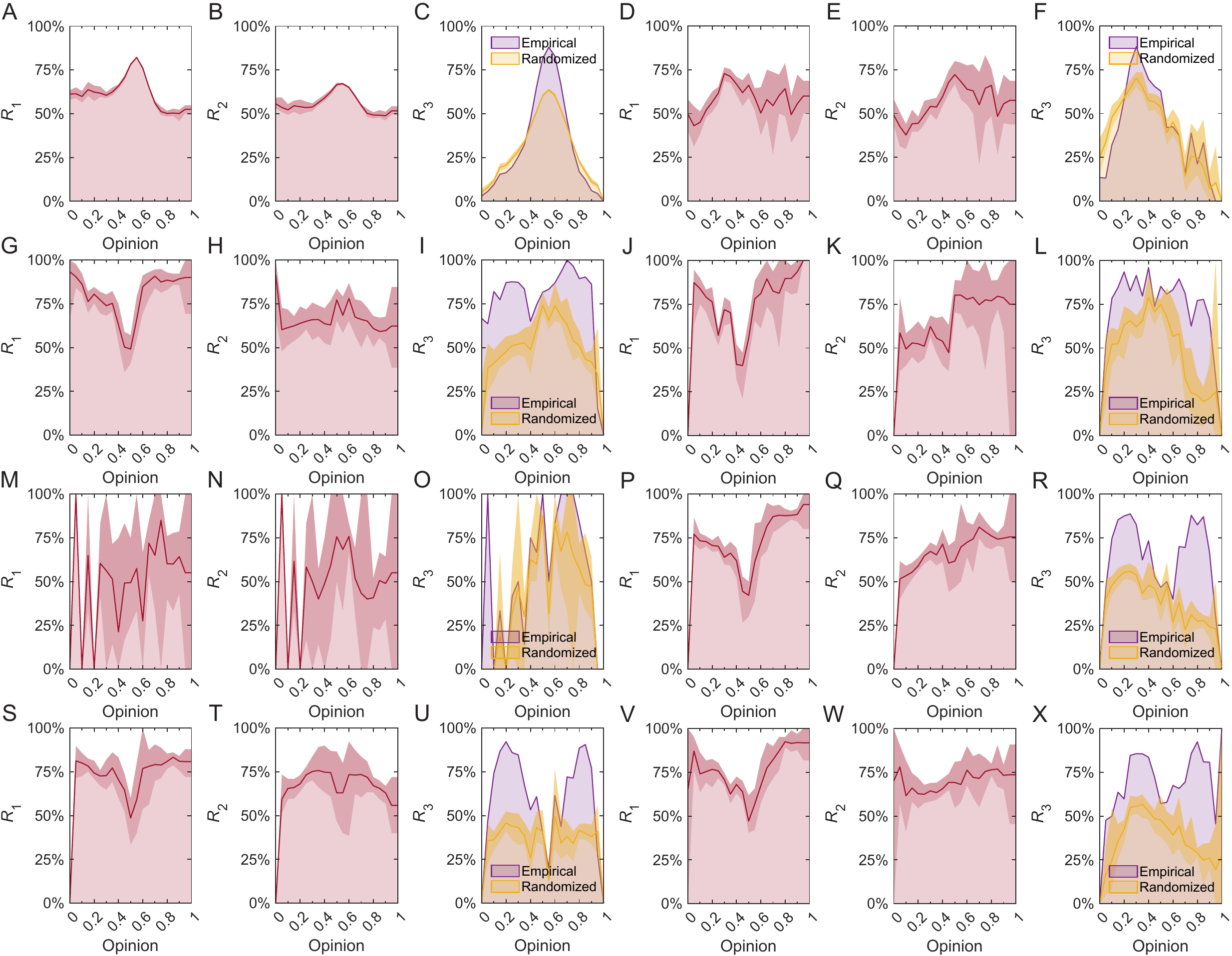}
	\caption{\textbf{{Rates of interval-based conditions} across partition-based opinion intervals in the first month of the Reddit-politics dataset and all Twitter datasets in the leader perspective.}
		Each row presents results for two datasets. Row 1: May 2018 of Reddit-politics, Twitter-Covid; Row 2: Twitter-contentious Abortion, Gun control; Row 3: Twitter-contentious Obamacare, US 2020 vice presidential debate; Row 4: Twitter-contentious US 2020 second presidential debate, US 2020 election day.
        Columns 1$\div$3: $R_1$, $R_2$, and $R_3$ for the first dataset in each row. Columns 4$\div$6: the same metrics for the second dataset.}
	\label{fig: case_1_fig_four_all}
\end{figure}

\begin{figure}[H]
	\centering
	\includegraphics[width=\linewidth]{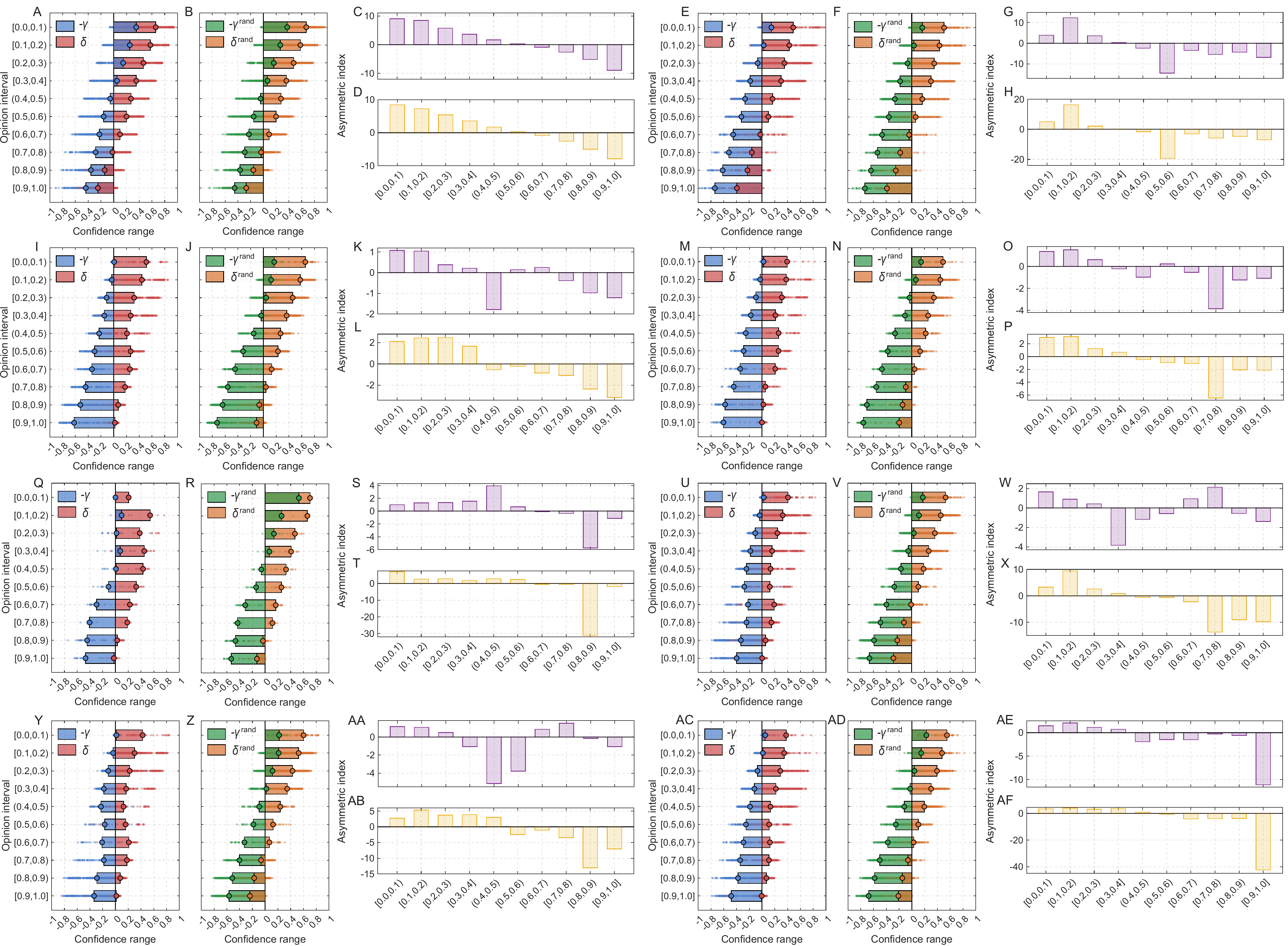}
	\caption{\textbf{Asymmetric {neighborhood spans} and opinion-dependent tolerance patterns in the first month of the Reddit-politics dataset and all Twitter datasets in the leader perspective.}
		Columns 1, 2, 4, 5: Distributions of negated left {offsets} ($-\gamma_i$) and right {offsets} ($\delta_i$) across different opinion intervals for May 2018 of Reddit-politics (A, B), Twitter-Covid (E, F), Twitter-contentious Abortion (I, J), Gun control (M, N), Obamacare (Q, R), US 2020 vice presidential debate (U, V), US 2020 second presidential debate (Y, Z), US 2020 election day (AC, AD).
		Columns 1 and 4 present empirical data, while Columns 2 and 5 show their corresponding range-based null models.
		Columns 3 and 6: Asymmetry index $s_i$ across different opinion intervals in empirical data (C, G, K, O, S, W, AA, AE), and the corresponding asymmetry index $s_i$ derived from range-based null models (D, H, L, P, T, X, AB, AF).}
	\label{fig: case_1_fig_add_LeftRightC_all}
\end{figure}

We also examined a leader perspective, defining neighbors as users who comment on (or follow) Ego, rather than those Ego actively seeks out. 
This is illustrated in Figs.~\ref{fig: case_1_fig_one_all}--\ref{fig: case_1_fig_add_LeftRightC_all} in this \textit{SI}.

The second column in Fig.~\ref{fig: case_1_fig_one_all} in this \textit{SI} shows the corresponding opinion gap distributions and contrasts them with the null model.
While these distributions confirm a general tendency toward homophily, structural differences between the empirical networks and the null baselines are much more evident when considering the confidence range and opinion deviation metrics.
For confidence range, shown in the first column in Fig.~\ref{fig: case_1_fig_two_all}, statistical tests confirm a significant difference ($p < 0.05$) in Reddit-politics ($d \approx 0.95$) and Twitter-Covid ($d \approx 0.25$). 
The opinion deviation metrics shown in the two rightmost columns in Fig.~\ref{fig: case_1_fig_two_all} further support this trend: in Reddit-politics, mean and maximal deviations are $d \approx 0.49$ and $d \approx 0.64$, while in Twitter-Covid, they are $d \approx 0.14$ and $d \approx 0.16$.
%Overall, these results suggest that while the leader perspective broadens somewhat the engagement compared to the follow perspective, interactions still tend to cluster within ideologically proximal groups, though the effect varies across datasets.
{Overall, these results indicate that the leader perspective yields a different but still bounded interaction structure, with dataset-dependent differences in the distributional statistics. In all cases, however, interactions still tend to cluster within opinion-proximal groups, though the strength of this tendency varies across datasets.}

\newpage

\bibliographystyle{unsrt}
{\small
\bibliography{biblo,bibliografia}
}